\pdfoutput=1
\newif\ifFullPaper
\FullPapertrue
\ifFullPaper
\documentclass[11pt]{article}
\usepackage{fullpage}
\else
\documentclass{llncs}
\fi

\usepackage{graphicx}
\usepackage{subfigure}
\usepackage{amssymb}
\usepackage{times}
\usepackage{url}
\usepackage{hyperref}
\usepackage[noend]{algorithmic}
\usepackage{cite}
\ifFullPaper
\newtheorem{lemma}{Lemma}
\newtheorem{theorem}[lemma]{Theorem}

\newcommand{\qed}{~\rule{6pt}{6pt}}
\newenvironment{proof}{\noindent{\bf Proof:~}}{\hspace*{\fill}\qed\bigskip}

\fi
\newcommand{\opt}{{\bf Opt}}
\newcommand{\R}{{\bf R}}
\newcommand{\Comment}[1]{\relax}        
\renewcommand{\subsection}[1]{\paragraph{#1.}}

\let\oldendproof\endproof
\def\endproof{\qed\oldendproof}

\ifFullPaper
\fi

\begin{document}

\title{On the Approximability of Geometric and Geographic
Generalization and the Min-Max Bin Covering Problem}
\ifFullPaper
\renewcommand{\thefootnote}{\fnsymbol{footnote}}
\author{
Wenliang Du\footnotemark[1]
\and
David Eppstein\footnotemark[2]
\and
Michael T.~Goodrich\footnotemark[2]
\and
George S. Lueker\footnotemark[2]
}
\date{}
\footnotetext[1]{Department of Electrical Engineering and
	   Computer Science, Syracuse University, 3-114 Sci-Tech Building.
	   Syracuse, NY 13244. \texttt{wedu(at)syr.edu}}
\footnotetext[2]{Department of Computer Science, 
	   University of California, Irvine. Irvine, CA USA 92697-3435.
   \texttt{\{eppstein,goodrich,lueker\}(at)ics.uci.edu}.}
\fi

\maketitle

\begin{abstract}
We study the problem of abstracting a table of data about individuals so
that no selection query can identify fewer than $k$ individuals.  
\ifFullPaper
As is
common in existing work on this \emph{$k$-anonymization} problem, the
means we investigate to perform this anonymization is to \emph{generalize}
values of quasi-identifying attributes into equivalence classes.
Since such data tables are intended for use in data mining, we consider
the natural optimization criterion of minimizing the maximum size of any
equivalence class, subject to the constraint that each is of size at
least $k$.  
\fi
We show that it is impossible to achieve arbitrarily good
polynomial-time approximations for a number of natural variations of
the generalization technique,
unless $P=NP$, even when the table has
only a single quasi-identifying attribute that represents a geographic
or unordered attribute:

\begin{itemize}
\item
\emph{Zip-codes}: 
nodes of a planar graph generalized into connected subgraphs
\item 
\emph{GPS coordinates}:
points in $\R^2$ generalized into non-overlapping rectangles
\item
\emph{Unordered data}:
text labels that can be grouped arbitrarily.
\end{itemize}

These hard single-attribute instances of generalization problems contrast
with the previously known NP-hard instances, which require the number
of attributes to be proportional to the number of individual records
(the rows of the table).  In addition to impossibility
results, we provide approximation algorithms for these difficult
single-attribute generalization problems, which, of course,
apply to multiple-attribute instances with one that is quasi-identifying.
\ifFullPaper
We show theoretically and experimentally that our approximation algorithms
can come reasonably close to optimal solutions.  
\fi
Incidentally, the generalization problem
for unordered data can be viewed
as a novel type of bin packing 
problem---\emph{min-max bin covering}---which may be of independent interest.
\end{abstract}

\section{Introduction}
\label{sec:introduction}

Data mining is an effective means for extracting useful
information from various data repositories, to highlight, for
example, health risks, political trends, consumer spending, 
or social networking.
In addition, some public institutions, such as the U.S.~Census Bureau,
have a mandate to publish data about 
U.S.~communities, so as to 
benefit socially-useful data mining.
Thus, there is a public interest in having data repositories
available for public study through data mining.
%
%
%

Unfortunately, fulfilling this public interest is complicated by the fact that
many databases contain confidential or personal information
about individuals. The publication of such information
is therefore constrained by laws and policies governing privacy protection.
For example, the U.S.~Census Bureau must limit its data releases to
those that reveal no information about any individual.
Thus, to allow the public to benefit from the 
knowledge that can be gained through
data mining, a privacy-protecting 
transformation should be performed on a database before
its publication. 

One of the greatest threats to privacy faced by database publication is a 
\emph{linking attack}~\cite{samarati98,samarati01}.
In this type of attack, an adversary who already knows partial 
identifying information about an individual 
(e.g., a name and address or zip-code)
is able to identify a record in another database
that belongs to this person.
A linking attack occurs, then, if an adversary
can ``link'' his prior identifying knowledge 
about an individual through a non-identifying attribute in another database.
Non-identifying attributes that can be subject to such linking attacks
are known as \emph{quasi-identifying} attributes.

To combat linking attacks,
several researchers~\cite{lefevre05,samarati01,samarati98,mw-cok-04,%
afkmptz-at-05,bkbl-ekuct-07,zyw-pekcd-05,ba-dpoka-05} have proposed
\emph{generalization} as a way of 
specifying a quantifiable privacy requirement for published databases.
The generalization approach is to group attribute values into equivalence
classes, and replace each individual attribute value with its class name. 
\ifFullPaper
Of course, 
we desire a generalization that is best for data mining purposes.
Thus, we add an additional constraint that we minimize the cost,
$C$, of our chosen generalization, where the
\emph{cost} of a set of equivalence classes
$\{E_1,E_2,\ldots,E_m\}$ is 
$ C = \sum_{i=1}^m c( E_i )$,
where $c$ is a cost function defined on equivalence classes
and the summation is defined in terms of 
either the standard ``$+$'' operator or the ``$\max$'' function.

The cost function $c$ should represent an optimization goal that is
expected to yield a table of transformed data that is the best for
data mining purposes, e.g., while preserving $k$-anonymity and/or
$l$-diversity~\cite{mkgv-ldpdk-07}.
Thus, we define the cost function $c$ 
in terms of a \emph{generalization error}, 
so, for an equivalence class $E=\{x_1,x_2,\ldots,x_q\}$,
\[
c(E) = \sum_{i=1}^q d(x_i,E),
\]
where $d$ is a measure of the difference between an element and its
equivalence class, and the summations are defined in terms of
either the standard ``$+$'' operator or the ``$\max$'' function.
For example, using $d(x_i,E)=1$, and taking ``$+$'' in the definition
of $C$ to be ``$\max$'',
amounts to a desire to minimize the maximum size of any equivalence class.
Alternatively,
using $d(x_i,E)=|E|-k$ and standard addition in the
definition of $C$ will quadratically penalize larger equivalence classes.
\fi
In this paper we focus on generalization methods
that try to minimize the maximum size of any
equivalence class, subject to lower bounds on the size of any
equivalence class. 
\ifFullPaper
This should also have the side effect of reducing 
the number of generalizations done, but we focus on minimizing the 
maximum equivalence
class here, as it leads to an interesting type of bin-packing problem, 
which may be of interest in its own right.

Since most prior work on $k$-anonymization algorithms has focused on numeric
or ordered data, we are interested in this paper on techniques that can be
applied on geographic and unordered data.
Such data commonly occurs in quasi-identifying attributes, but 
such attributes seem harder to generalize to achieve $k$-anonymity. Thus, we
are interested in the degree to which one can approximate the optimal way of
generalizing geographic and unordered data, using natural generalization
schemes.
\fi

\subsection{Related Prior Results}
\label{sec:related_work}

The concept of $k$-anonymity~\cite{samarati98,samarati01}, 
although not a complete solution to linking attacks, 
is often an important component of such solutions.
In this application of generalization,
the equivalence classes are chosen to ensure that each combination of replacement attributes that occurs in the generalized database occurs in at least $k$ of the records.
Several researchers have explored heuristics, extensions, and adaptations
for $k$-anonymization
(e.g., see~\cite{lefevre05,afkmptz-at-05,bkbl-ekuct-07,zyw-pekcd-05,ba-dpoka-05,wf-asr-06}).

\ifFullPaper
As mentioned above,
generalization has become a popular way
of altering a database (represented as a table) 
so that it satisfies the $k$-anonymity requirement,
by combining attribute values into equivalence classes.
To guide this ``combining'' process for a particular attribute, 
a generalization hierarchy (or concept hierarchy) is often specified, 
which is either derived from an ordering on the data or itself defines
an ordering on the data. 

Unfortunately, there is no obvious tree hierarchy 
for geographic and unordered data.
So, for unordered data, several researchers have introduced heuristics for
deriving hierarchies that can then be used for generalization.
Security properties of the
randomization schemes and privacy-preserving data mining in general
are studied by Kargupta {\it et al.}~\cite{KarDaWaSi03}, 
Kantarcioglu {\it et al.}~\cite{KanJinClif04}, and Huang {\it et al.}~\cite{HuangDu05}.
(See also Fung~\cite{Fung07thesis}.)
Wang {\it et al.}~\cite{WangYuCha2004}
used an iterative bottom-up heuristic to generalize data.
\fi

The use of heuristics, rather than exact algorithms, for performing 
generalization is
motivated by claims that $k$-anonymization-based generalization is NP-hard.
Meyerson and Williams~\cite{mw-cok-04}
assume that an input dataset has been processed into a database or table in which identical records from the original dataset have been aggregated into a
single row of the table, with a count representing its frequency.
They then show that if the number of aggregated rows is $n$ and the number of attributes (table columns) is at least $3n$,
then generalization for $k$-anonymization is
NP-hard. Unfortunately, their proof does not show that
generalization is NP-hard in the strong sense: the difficult instances generated by their reduction have frequency counts that are large binary numbers, rather than being representable in unary. Therefore, their result doesn't
actually apply to the original $k$-anonymization problem.
Aggarwal {\it et al.}~\cite{afkmptz-at-05} address this 
deficiency, showing that $k$-anonymization is NP-hard
in the strong sense 
for datasets with at least $n/3$ quasi-identifying attributes.
Their proof uses cell suppression instead of generalization, but Byun
{\it et al.}~\cite{bkbl-ekuct-07} show that the proof can be extended to
generalization.
As in the other two NP-hardness proofs, Byun
{\it et al.} require that the number of quasi-identifying attributes be proportional to the number
of records, which is typically not the case.
Park and Shim~\cite{ps-aaka-07} present an NP-hardness proof
for a version of $k$-anonymization involving cell 
suppression in place of generalization, and
Wong {\it et al.}~\cite{wlfw-akekm-06} show an anonymity problem
they call $(\alpha,k)$-anonymity to be NP-hard.

Khanna {\it et al.}~\cite{kmp-oartp-98} study a problem, RTILE, which is closely related to generalization of geographic data. RTILE involves tiling an
$n\times n$ integer grid with at most $p$ rectangles so as to
minimize the maximum weight of any rectangle.
They show that no polynomial-time approximation algorithm can achieve an
approximation ratio for RTILE of better than $1.25$ unless P$=$NP. 
Unlike $k$-anonymization, however, this problem does not constrain the 
minimum weight of a selected rectangle.
\ifFullPaper
Aggarwal~\cite{a-kcd-05} studies the problem of generalizing multidimensional
data using axis-aligned rectangles using probabilistic clustering techniques,
and Hore {\it et al.}~\cite{hjm-fappd-07} study a heuristic based on 
the use of kd-tree partitioning and a search strategy optimized through 
the use of priority queues.
Neither of these papers gives provable approximation ratios,
however.
\fi

\subsection{Our Results}
In this paper, we study instances of $k$-anonymization-based generalization in which there is only a single
quasi-identifying attribute, containing geographic or unordered data. 
In particular, we focus on the following attribute types:
\begin{itemize}
\item
\emph{Zip-codes:}
nodes of a planar graph generalized into connected subgraphs
\item 
\emph{GPS coordinates:}
points in $\R^2$ generalized into non-overlapping rectangles
\item
\emph{Unordered data}:
text labels that can be grouped arbitrarily (e.g., disease names).
\end{itemize}
We show that even in these simple instances, $k$-anonymization-based generalization is NP-complete in the strong sense.
Moreover, it is impossible to approximate these problems to within
$(1+\epsilon)$ of optimal,
where $\epsilon>0$ is an arbitrary fixed constant, unless $P=NP$.
These results hold \emph{a fortiori} for instances with multiple quasi-identifying attributes of these types, and they greatly strengthen previous NP-hardness results
which require unrealistically large numbers of attributes.
Nevertheless, we provide a number of efficient approximation
algorithms and 
\ifFullPaper
we show, both in terms of their worst-case 
approximation performance and also in terms of their empirical
performance on real-world data sets, 
\fi
that they achieve good
approximation ratios.  Our approximation bounds for the zip-codes problem
require that the graph has sufficiently strong connectivity to guarantee
a sufficiently low-degree spanning tree.


The intent of this paper is not to argue that 
single-attribute generalization
is a typical application of privacy protection.
Indeed, most real-world anonymization applications will have dozens 
of attributes whose privacy concerns vary from 
hypersensitive to benign.
Moreover, the very notion of 
$k$-anonymization has been shown to be insufficient to protect against all types of linking attack, and has been extended recently in various 
ways to address some of those concerns (e.g., see~\cite{dt-ckasi-08,mkgv-ldpdk-07,wlfw-akekm-06});
some work also argues against any approach similar to $k$-anonymization
\cite{Dwork:AdOmnia}.
We do not attempt to address this issue here.
Rather, our results should be viewed as showing that even the simplest forms of $k$-anonymization-based generalization
are difficult but can be approximated. We anticipate that similar results may hold for its generalizations
and extensions as well.

In addition, 
from an algorithmic perspective, our study of 
$k$-anony\-mi\-za\-tion-based generalization
has uncovered a new kind of bin-packing problem (e.g.,
see~\cite{cgj-aabp-97}),
which we call \emph{Min-Max Bin Covering}.
In this variation, we are given a collection of items and a nominal bin
capacity, $k$, and we wish to distribute the items to bins so that each bin
has total weight at least $k$ while minimizing the maximum weight of any bin.
This problem may be of independent interest in the algorithms
research community.

\ifFullPaper
Incidentally, our proof that $k$-anonymization is NP-hard for points in the
plane can be easily adopted to show that the RTILE problem, studied
by Khanna {\it et al.}~\cite{kmp-oartp-98}, cannot be approximated 
in polynomial time by a factor better than $1.33$, unless P$=$NP, 
which improves the previous non-approximability 
bound of $1.25$.
\fi

\section{Zip-code Data}
The first type of quasi-identifying information we consider is that of zip-codes, or analogous numeric codes for other geographic regions.
Suppose we are given a database consisting of $n$ records, each of which
contains a single quasi-identifying
attribute that is itself a zip-code.
A common approach in previous papers 
using generalization for zip-code data
(e.g., see~\cite{bkbl-ekuct-07,zyw-pekcd-05})
is to generalize consecutive zip-codes.
That is, these papers view zip-codes as character strings or
integers and generalize based on this data type.
Unfortunately, \ifFullPaper as is illustrated in Figure~\ref{fig:zips},\fi
when zip-codes are viewed as numbers or strings,
geographic adjacency information can be lost or misleading: consecutive zip codes may be far apart geographically, and geographically close zip codes may be numerically far, leading to generalizations that have poor quality for data mining applications.

\ifFullPaper
\begin{figure}[hbt]
\centering\includegraphics[width=3.3in]{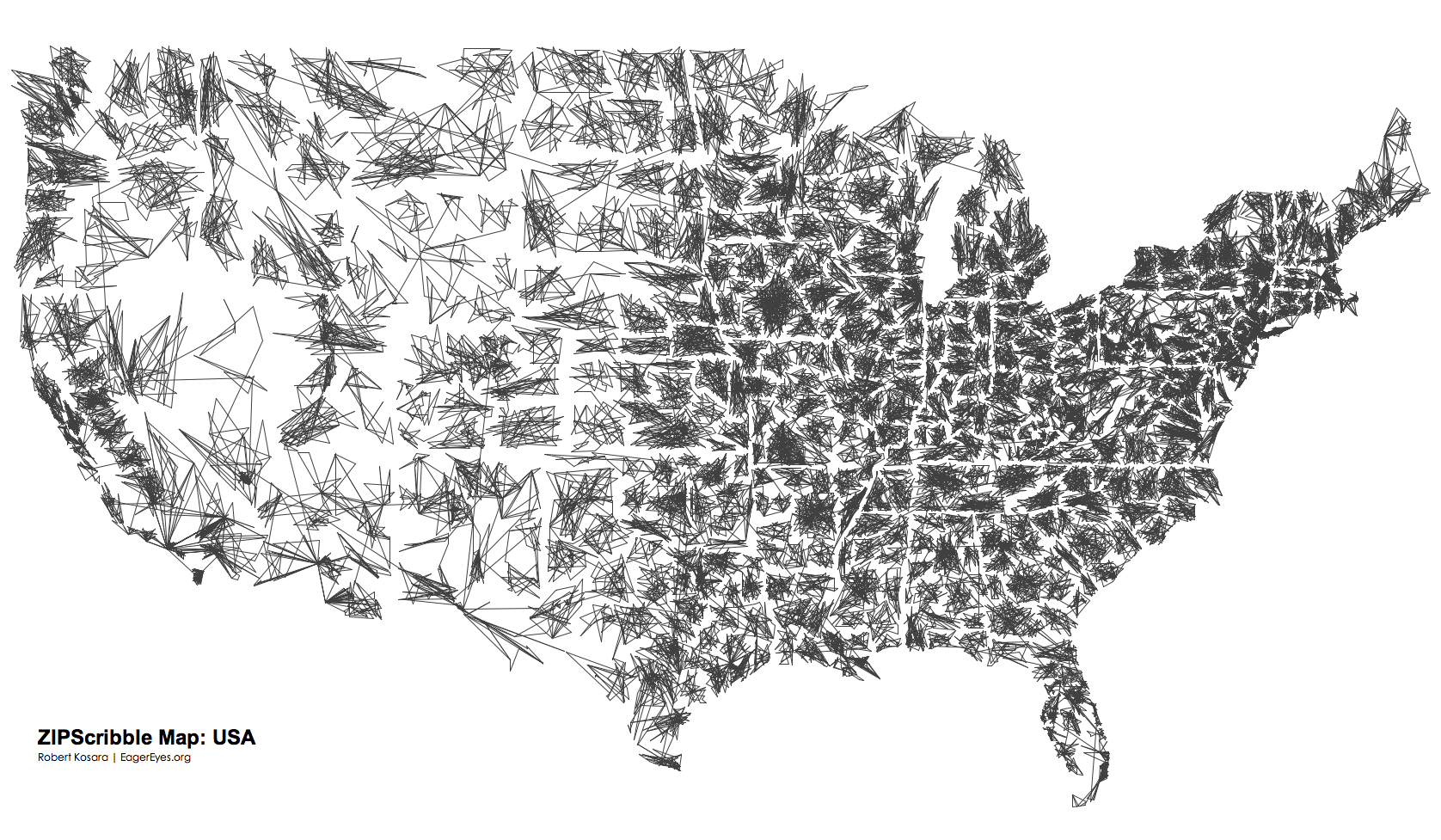}
\caption{The US ZIPScribble Map, which connects consecutive
zipcodes. Note the lack of proximity preservation in the West
and the artificial separations in the East.
(From {http://eagereyes.org/Applications/ZIPScribleMap.html}.)}
\label{fig:zips}
\end{figure}
\fi

We desire a generalization algorithm for zip-codes that preserves
geographic adjacency.
Formally, we assume each zip-code is
the name of a node in a planar graph, $G$.
The most natural generalization
in this case is to group 
nodes of $G$ into equivalence classes that are connected subgraphs.
This is motivated, in the zip-code case, by a desire to
group adjacent regions in a country, which would naturally have more
likelihood to be correlated according to factors desired
as outcomes from data mining, such as health or buying
trends.
So the optimization problem we investigate in this section is one in
which we are given a planar graph, $G$, with non-negative integer weights 
on its nodes (representing the number of records for each node), 
and we wish to partition $G$ into connected subgraphs
so that the maximum weight of any subgraph is minimized subject to
the constraint that each has weight at least $k$.

\subsection{Generalization for Zip-codes is Hard}
Converting this to a decision problem, we can add a parameter
$K$ and ask if there exists a partition into connected subgraphs such that the weight of each subgraph in $G$ is at least $k$ and
at most $K$.
In this section, we show that this problem is NP-complete even if the
weights are all equal to $1$ and $k=3$.
Our proof is based on a simple reduction that sets $K=3$, so as to
provide a reduction from the following problem:
\begin{trivlist}
\item
\textbf{3-Regular Planar Partition into Paths of Length 2 (3PPPL2)}:
Given a 3-regular planar graph $G$, can $G$ be
partitioned into paths of length 2? 
That is, is there a spanning forest for $G$ such that each connected
component is a path of length 2?
\end{trivlist}

This problem is a special case of the problem, 
``Partition into Paths of Length-2 (PPL2)'', whose NP-completeness is
included as an exercise in Garey and Johnson~\cite{gj-cigtn-79}.
Like PPL2, 3PPPL2 is easily shown to be in NP.
To show that 3PPPL2 is NP-hard, we provide a reduction from
the 3-dimensional matching (3DM) problem:
\begin{trivlist}
\item
\textbf{3-Dimensional Matching (3DM)}:
Given three sets $X$, $Y$, and $Z$, each of size~$n$, and a
set of triples $\{(x_1,y_1,z_1),\ldots,(x_m,y_m,z_m)\}$, is
there a subset $S$ of $n$ triples such that each element in $X$, $Y$,
and $Z$ is contained in exactly one of the triples?
\end{trivlist}

\begin{figure}[hb!]
\begin{tabular}{c@{\hspace*{0.75in}}c}
\includegraphics[scale=.2]{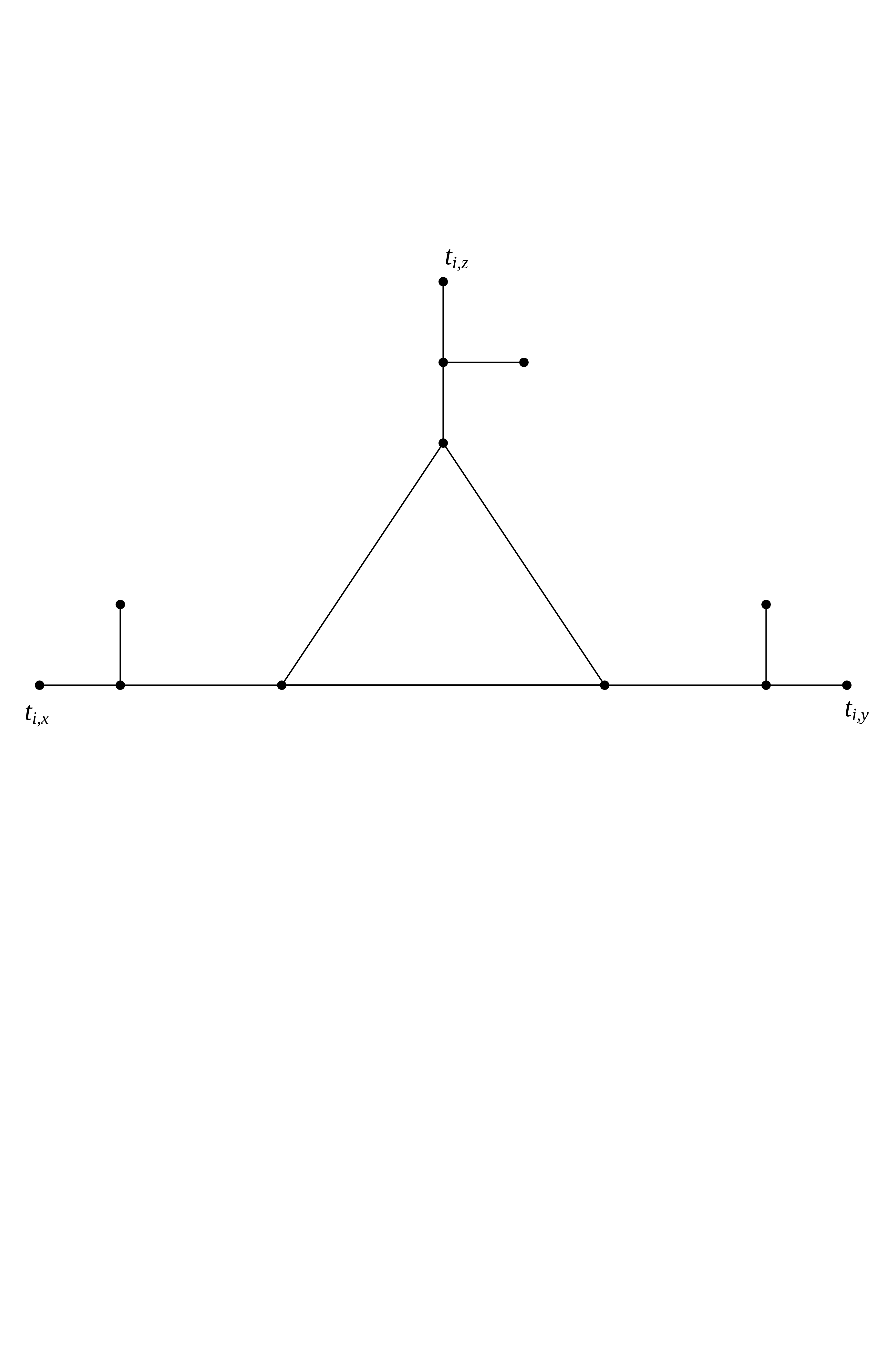} &
\includegraphics[scale=.2]{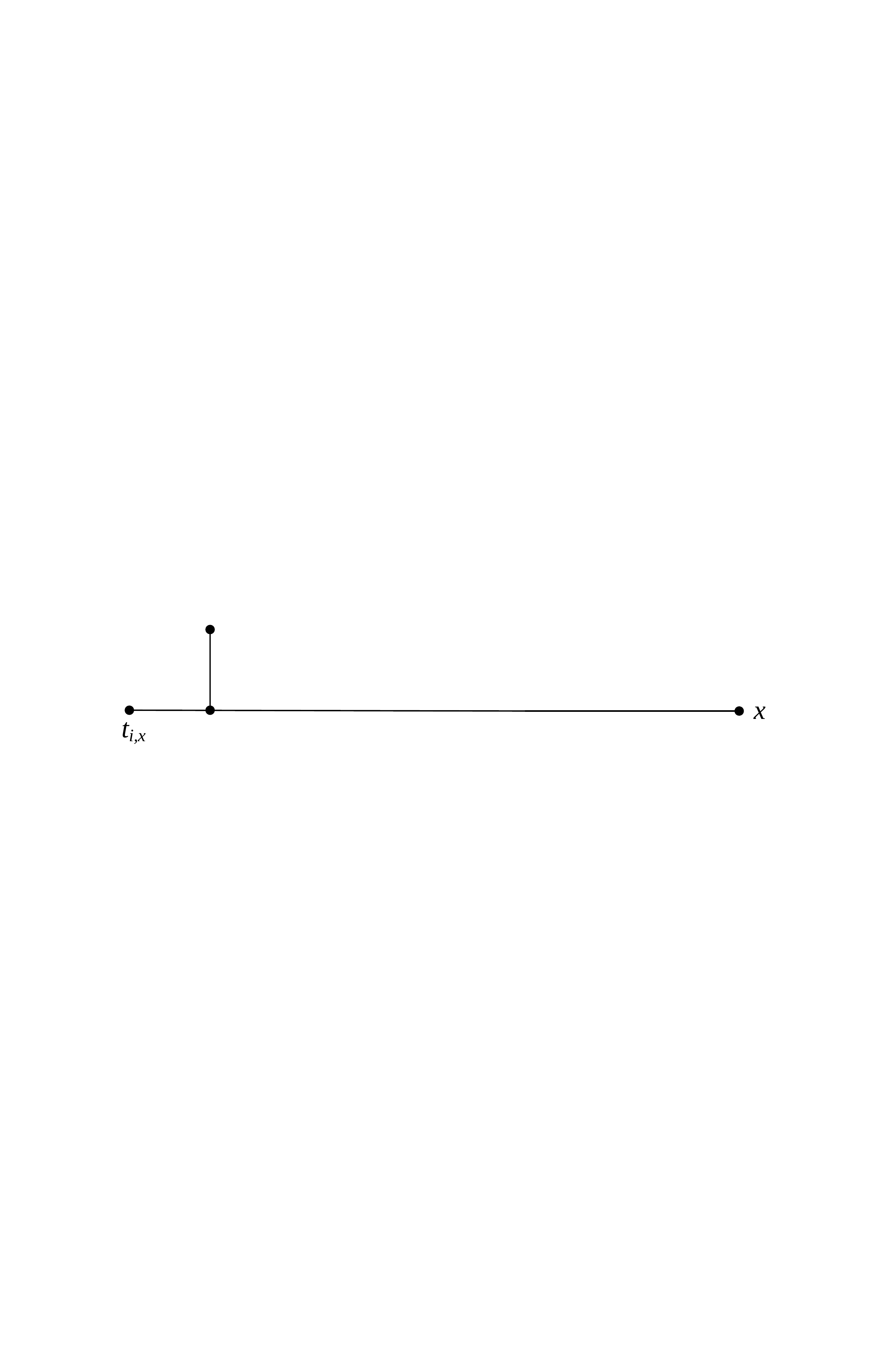} \\[-1.55in]
(a) & (b)
\end{tabular}
\vspace*{-8pt}
\caption{\label{fig-tuple} Gadgets for reducing 3DM to PPL2.
(a) the tuple gadget; (b) the connector.}
\end{figure}

Suppose we are given an instance of 3DM.
We create a vertex for each element in $X$, $Y$, and $Z$.
For each tuple, $(x_i,y_i,z_i)$, we create a tuple subgraph gadget as
shown in Figure~\ref{fig-tuple}a, with nodes 
$t_{i,x}$,
$t_{i,y}$,
and
$t_{i,z}$, which correspond to the representatives 
$x_i$, 
$y_i$, 
and
$z_i$, respectively, in the tuple.
We then connect each $t_{i,x}$, $t_{i,y}$ and $t_{i,z}$
vertex to the corresponding element vertex from $X$, $Y$, and $Z$,
respectively, using the connector gadget in Figure~\ref{fig-tuple}b.

\begin{figure}[hb!]
\vspace*{-1.1in}
\begin{tabular}{c@{\hspace*{0.75in}}c}
\includegraphics[scale=0.19]{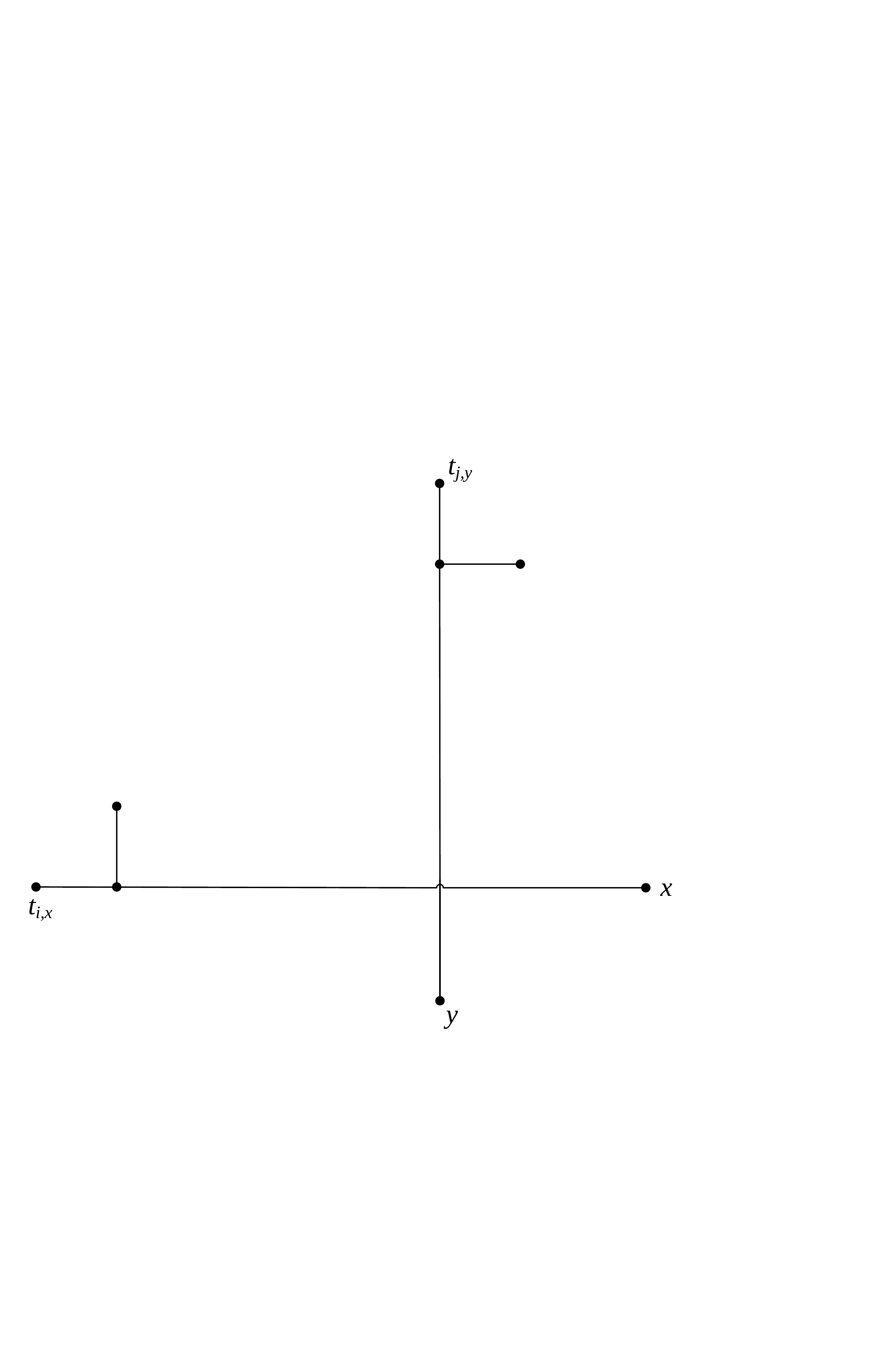} &
\includegraphics[scale=0.19]{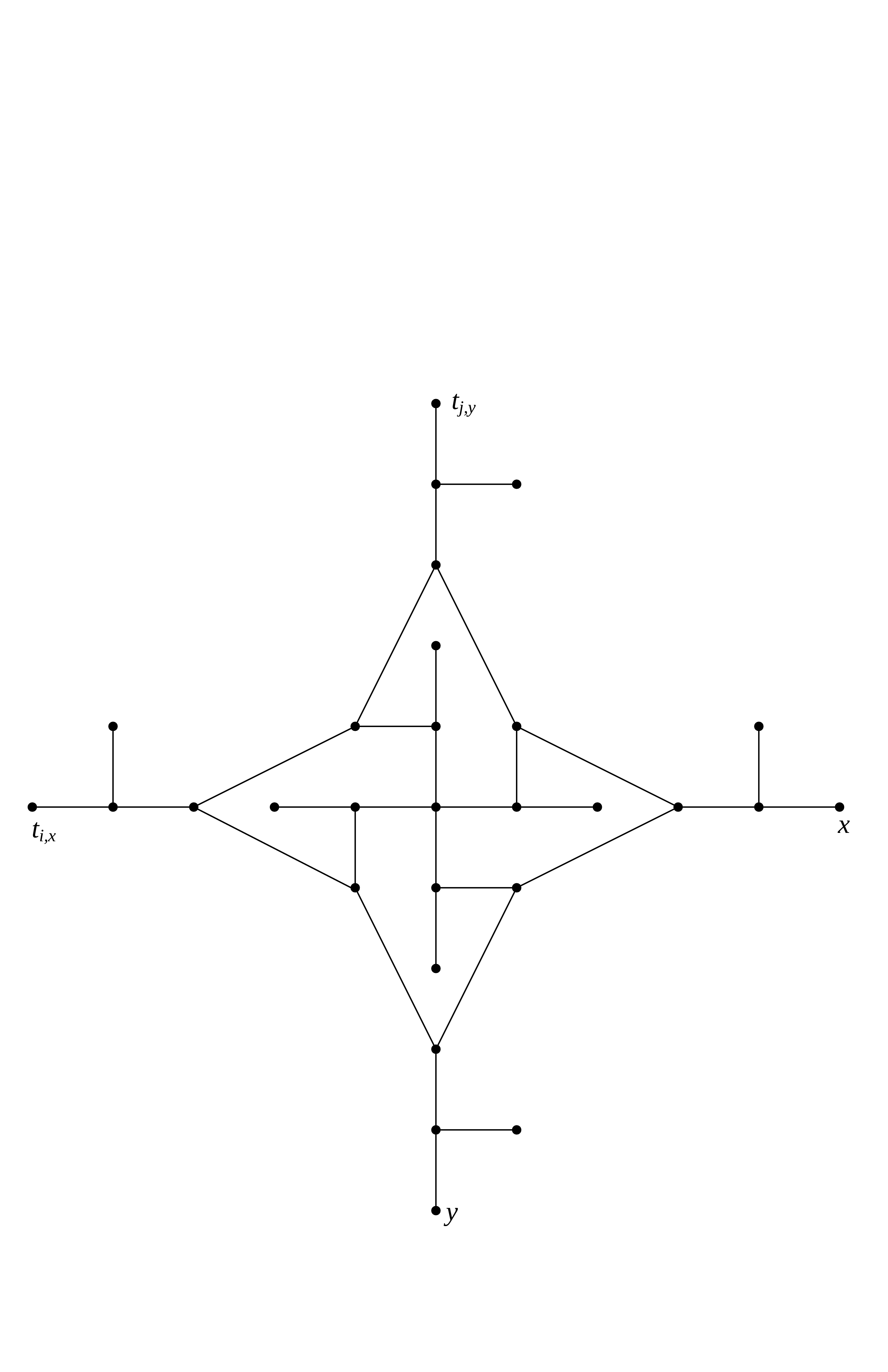} \\[-0.3in]
(a) & 
(b) 
\end{tabular}
\vspace*{-8pt}
\caption{\label{fig-cross}
  Dealing with edge crossings. (a) a connector crossing; (b) the
  cross-over gadget.}
\end{figure}

\ifFullPaper
\begin{figure}[hbt]
\vspace*{-2in}
\begin{center}
\begin{tabular}{c@{\hspace*{0.1in}}c}
\hspace*{-1.5in}\includegraphics[scale=0.4]{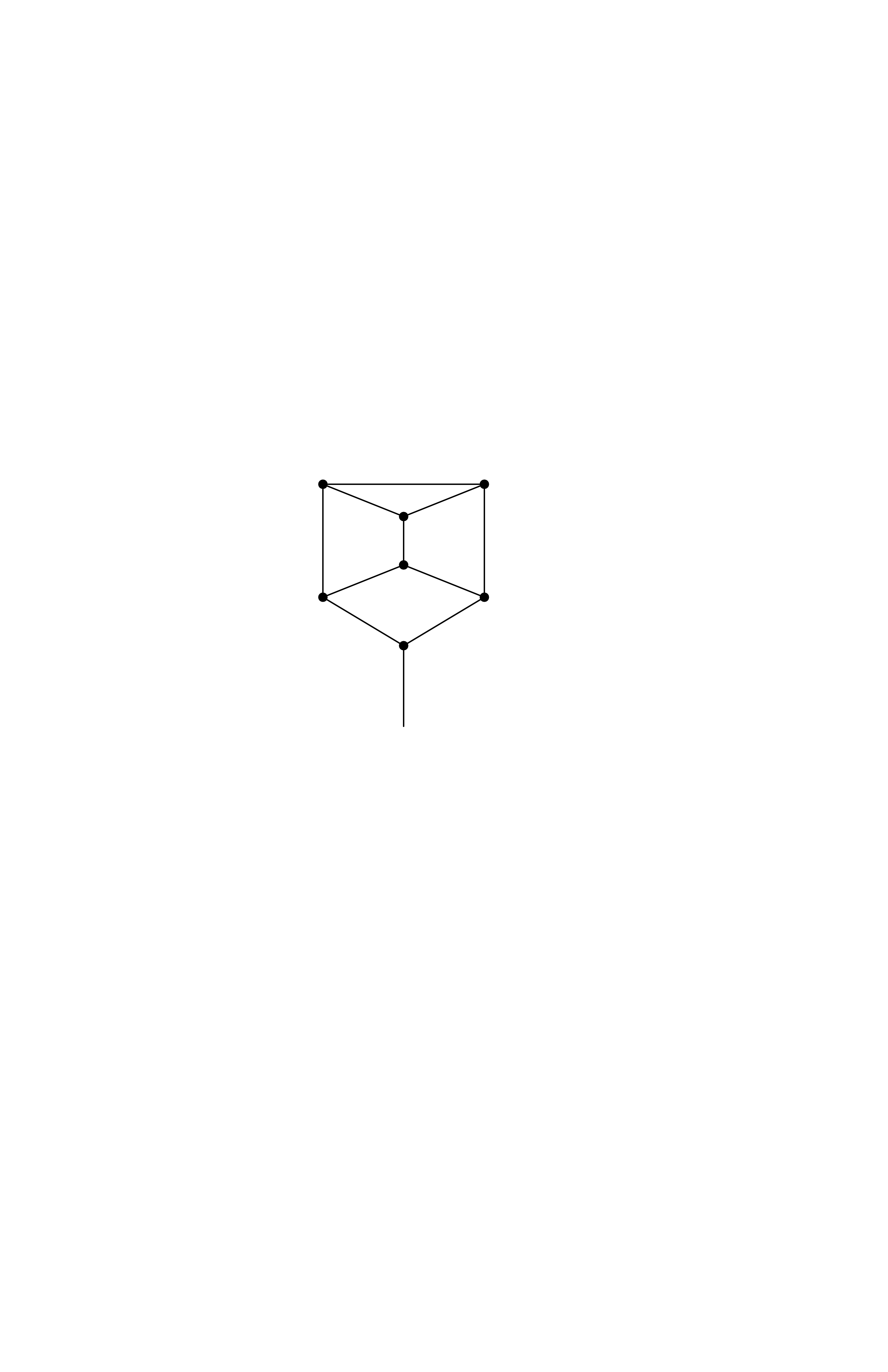} &
\hspace*{-2.5in}\includegraphics[scale=0.4]{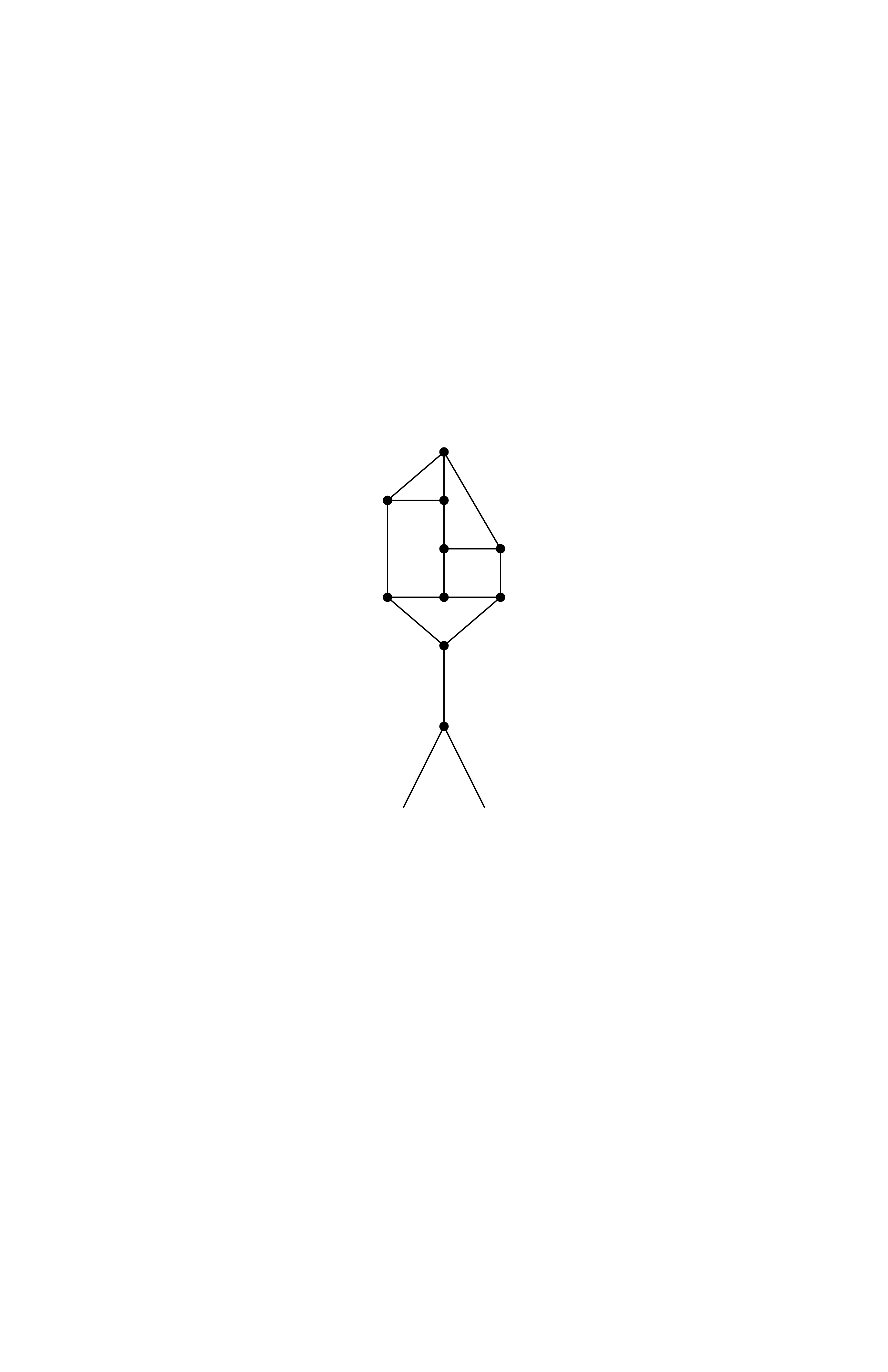} \\[-2.75in]
\hspace*{-1.75in}(a) &
\hspace*{-2.5in}(b) 
\end{tabular}
\end{center}
\caption{\label{fig-cap}
  Augmentations to achieve 3-regularity. 
  (a) a ``cap'' to add to a degree-1 vertex;
  (b) a ``cap'' to add to a degree-2 vertex.}
\end{figure}
\fi

This construction is, in fact, a version of 
the well-known folklore reduction from
3DM to PPL2, which solves an exercise in Garey and Johnson~\cite{gj-cigtn-79}. 
Note, for example, that the vertices in the triangle in the tuple gadget 
must all three be completely included in a single group or must all
be in separate groups.
If they are all included, then grouping the degree-1 vertices
requires that the corresponding $x$, $y$, and $z$ elements must all
be included in a group with the degree-1 vertex on the connector.
If they are all not included, then the corresponding $x$, $y$, and
$z$ elements must be excluded from a group in this set of gadgets.

Continuing the reduction to an instance of 3PPPL2, we make a series
of transformations.
The first is to embed the graph in the plane in such a way that the only crossings occur in connector gadgets. We then take each crossing of a connector, as shown in
Figure~\ref{fig-cross}a, and replace it with the cross-over gadget
shown in Figure~\ref{fig-cross}b.

\ifFullPaper
There are four symmetric ways this gadget can be partitioned into
paths of length 2, two of which are shown in Figures~\ref{fig-cross2}a
and \ref{fig-cross2}b.
\fi
Note that the four ways correspond to the four possible ways that
connector ``parity'' can be transmitted and that they correctly
perform a cross-over of these two parities. In particular, note that
it is impossible for opposite connectors to have the same parity
in any partition into paths of length 2.
Thus, replacing each crossing with a cross-over gadget completes a
reduction of 3DM to planar partition in paths of length 2.

\ifFullPaper
\begin{figure}[hbt]
\vspace*{-0.7in}
\begin{tabular}{c@{\hspace*{0.75in}}c}
\includegraphics[scale=0.24]{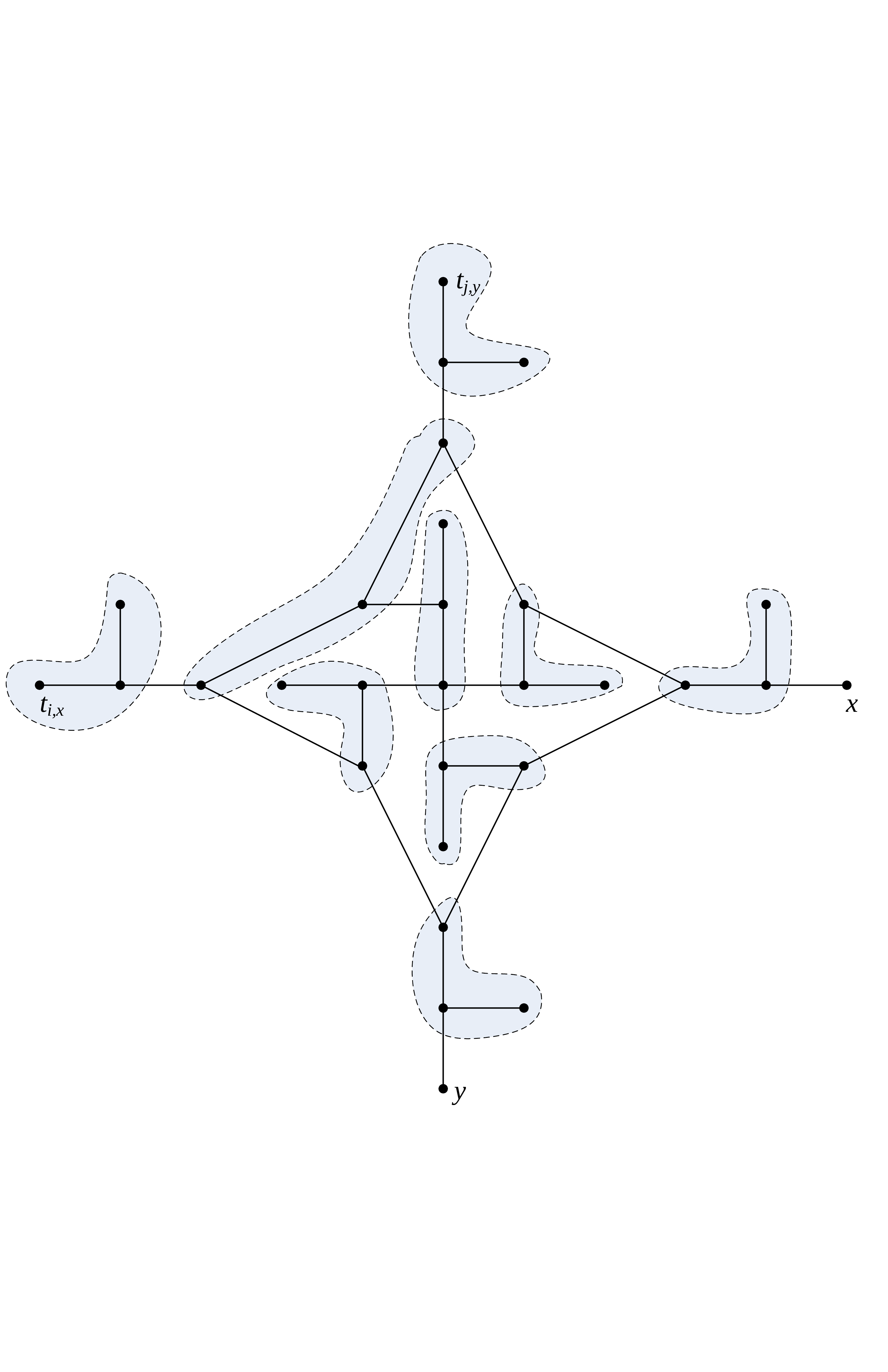} &
\includegraphics[scale=0.24]{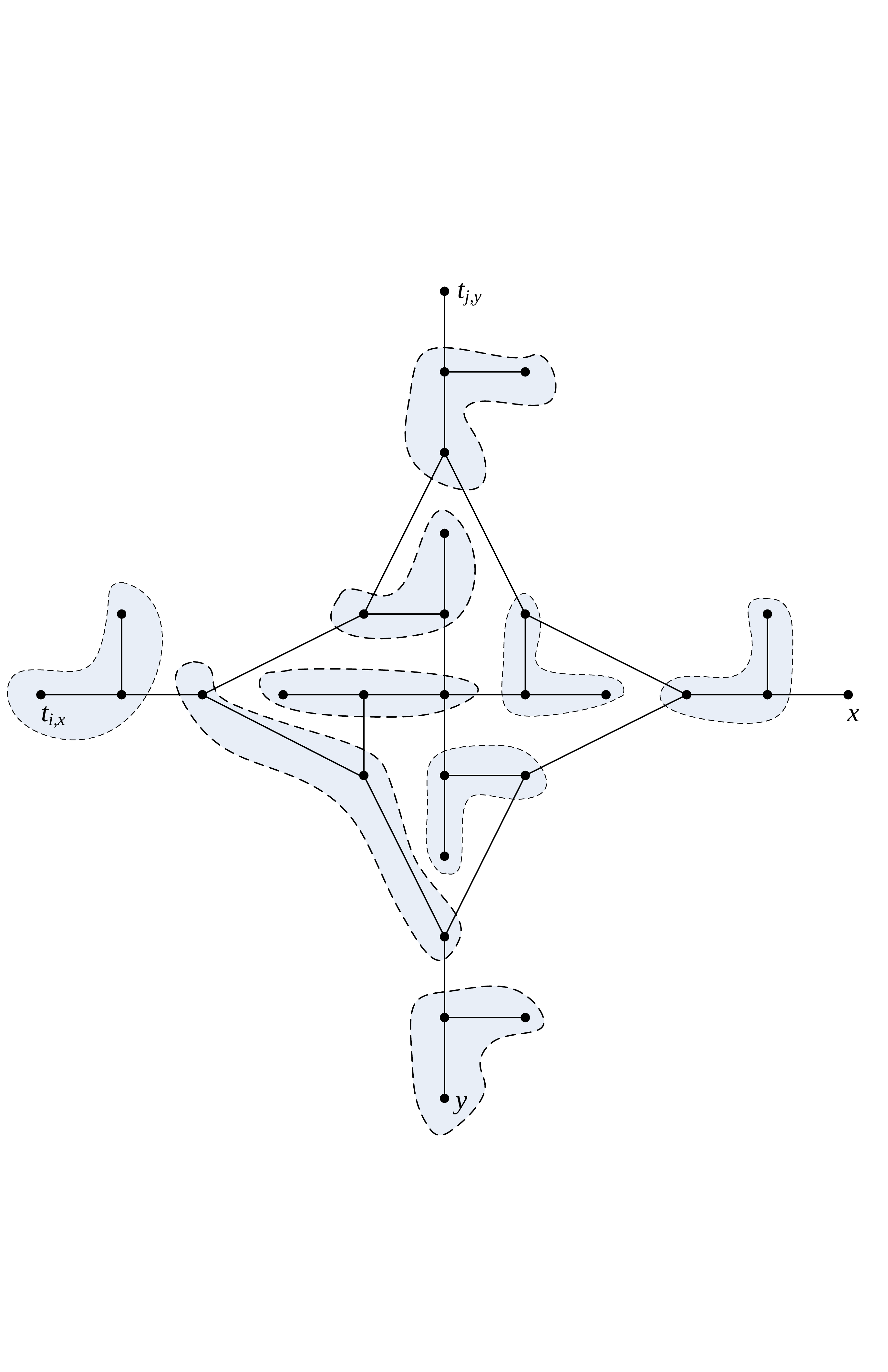} \\[-0.75in]
(a) & (b) 
\end{tabular}
\vspace*{-8pt}
\caption{\label{fig-cross2}
  Dealing with crossings. 
  (a) one possible partition of the cross-over
  gadget; (b) another one of the four possible partitions.}
\end{figure}
\fi

Next, note that all vertices of the planar graph are degree-3 or less
except for the ``choice'' vertices at the center of cross-over gadgets and
possibly some nodes corresponding to elements of $X$, $Y$, and $Z$.
For each of these, we note that all the edges incident on such nodes
are connectors.
We therefore replace each vertex of degree-4 or higher with three
connector gadgets that connect the original vertex to three binary trees
whose respective edges are all connector gadgets.
This allows us to ``fan out'' the choice semantics of the original
vertex while exclusively using degree-3 vertices.
To complete the reduction, we perform additional simple transformations 
to the planar graph to make it 3-regular.
\ifFullPaper
In particular, we add to each degree-1 vertex the ``cap'' 
gadget shown in Figure~\ref{fig-cap}a.
Likewise, we add to each degree-2 vertex the cap shown
in Figure~\ref{fig-cap}b.
Note that in both cases, these subgraphs must be partitioned into
paths of length 2 that do not extend outside the subgraph. Thus,
adding these subgraphs to the original graph does not alter a
partition into paths of length 2 for the original graph.
\fi
This completes the reduction of 3DM to 3PPPL2.

\ifFullPaper
\subsection{An Approximation Algorithm for Zip-codes}
In this section we provide an approximation algorithm
for $k$-anonymization of zip-codes.
Suppose, therefore, that we are given a connected planar graph $G$
with non-negative integer vertex weights, and we wish to 
partition $G$ into connected subgraphs of weight at least $k$, while minimizing
the maximum weight of any subgraph.

We start by forming a low-degree spanning tree $T$ of $G$; let $d$ be the degree of $T$. We note that 3-connected planar graphs are guaranteed to have a spanning tree of degree three~\cite{Bar-CJM-66}, giving $d=3$, while Tutte~\cite{Tut-TAMS-56} proved that 4-connected planar graphs are always Hamiltonian, giving $d=2$; see~\cite{ChiNis-Algs-89,ChrKan-IJCGA-97,Str-PhD-97} for algorithms to construct $T$ efficiently in these cases. We then find an edge $e$ such that removing $e$ from $T$ leaves two trees $T_1$ and $T_2$, both of weight at least $k$, with the weight of $T_1$ as small as possible. If such an edge exists, we form one connected subgraph from $T_1$ and continue to partition the remaining low-degree tree $T_2$ in the same fashion; otherwise, we form a single connected subgraph from all of $T$.

Let $\kappa=\max(k,x_1,x_2,\dots)$ where $x_i$ are the individual item sizes; clearly, the optimal cost of any solution is at least $\kappa$.

\begin{lemma}
If the algorithm outlined above cannot find any edge $e$ to split $T$, then the cost of $T$ is at most $\kappa+d(k-1)$.
\end{lemma}

\begin{proof}
Orient each edge $e$ of $T$ from the smaller weight subtree formed by
splitting at $e$ to the larger weight subtree; if a tie occurs break
it arbitrarily.  Then $T$ must have a unique vertex $v$ at which all
edges are oriented inwards. The weight of $v$ is at most $\kappa$, and
it is adjacent to at most $d$ subtrees each of which has weight at
most $k-1$ (or else the edge connecting to that subtree would have
been oriented outwards) so the total weight of the tree is at most
$\kappa+d(k-1)$ as claimed.
\end{proof}

\begin{lemma}
If the algorithm above splits tree $T$ into two subtrees $T_1$ and $T_2$, then the cost of $T_1$ is at most $\kappa+(d-1)(k-1)$.
\end{lemma}

\begin{proof}
Let $v$ be the node in $T_1$ adjacent to the cut edge $e$. Then the weight of $v$ is at most $\kappa$, and $v$ is adjacent to at most $d-1$ subtrees of $T_1$ (because it is also adjacent to $e$). Each of these subtrees has weight at most $k-1$, or else we would have cut one of the edges connecting to them in preference to the chosen edge $e$. Therefore, the total weight is at most $\kappa+(d-1)(k-1)$ as claimed.
\end{proof}

\fi

\label{sect:zipcodes}
\begin{theorem} \label{thm:zipcodes}
There is a polynomial-time 
approximation algorithm for $k$-anonymization on planar
graphs that guarantees an approximation ratio of $4$ for $3$-connected planar graphs and
$3$ for $4$-connected planar graphs.
It is not possible for a polynomial-time algorithm to achieve an
approximation ratio better than $1.33$, even for 3-regular planar
graphs, unless P$=$NP.
\end{theorem}

\ifFullPaper
\begin{proof}
For 3-connected planar graphs, $d=3$, and the lemmas above show that our algorithm produces a solution with quality at most $\kappa+3(k-1)\le 4\kappa\le 4\,{\rm OPT}$. Similarly, for 4-connected planar graphs, $d=2$ and the lemmas above show that our algorithm produces a solution with quality at most $\kappa+2(k-1)\le 3\kappa\le 3\,{\rm OPT}$.

The inapproximability result follows from the NP-completeness result in the main text of the paper, as the graph resulting from that reduction either has a partition into 3-vertex connected subgraphs or some subgraph requires four or more vertices.
\end{proof}
\fi

\section{GPS-Coordinate Data}
Next we treat geographic data that is given as geographic coordinates rather than having already been generalized to zip-codes.
Suppose we are given a table consisting of $n$ records, each of
which contains a single quasi-identifying
attribute that is itself a GPS coordinate,
that is, a point $(x,y)$ in the plane.
\ifFullPaper
For example, the
quasi-identifying attribute could be the GPS coordinate of a home 
or elementary school.
\fi
Suppose further that we wish to generalize such sets of points using
axis-aligned rectangles.

\subsection{Generalizing GPS-Coordinates is Hard}
Converting this to a decision problem, we can add a parameter
$K$ and ask whether there exists a partition of the plane into rectangles such that the weight of the input points within each rectangle is at least $k$ and
at most $K$.
\ifFullPaper
In this section, we show that this problem is NP-complete even when we set $k$ 
and $K$ equal to~3.
Our proof is based on a simple reduction from 3-dimensional matching (3DM).
\fi

\ifFullPaper
Suppose we are given an instance of 3DM.
We first reduce this instance to a rectangular $k$-anonymization
problem inside a rectilinear polygon with holes; we show later
how to replace the edges of the polygon by points.
We begin by creating a separate point for each 
element in $X$, $Y$, and $Z$.
For each tuple, $(x_i,y_i,z_i)$, we create a tuple gadget as
shown in Figure~\ref{fig-box-tuple}a; the three points
in the interior must all be contained in a single rectangle or each
of them must be in a separate rectangle joining the two points that
sit in the ``doorway'' of a ``corridor.''
Inside the corridor, we alternatively place singleton points and
pairs of points, placing each singleton or pair at a corner of the corridor,
so that the only way to cover three points within the corridor is to use both points of a pair and one nearby singleton point; thus, any covering of all of the points of the corridor by rectangles containing exactly three points must preserve the parity of the connections at the adjacent doorways.
For each tuple $(x_i,y_i,z_i)$, we route the corridors from our tuple
gadget to each of the points $x_i$, $y_i$, and $z_i$, so that the
corridors for any point, such as $x_i$, meet in a chooser gadget 
as shown in Figure~\ref{fig-box-tuple}b.
Note: if the degree of a point grows to more than three, we can
fan-in the corridors in binary trees whose internal nodes are
represented with chooser gadgets.
Of course, some corridors may cross each other in this drawing, in
which case we replace each corridor crossing with the crossing gadget
shown in Figure~\ref{fig-box-tuple}c.

\begin{figure}[htb]
\vspace*{-3.4in}
\begin{tabular}{ccc}
\hspace*{-0.5in}\includegraphics[scale=.5]{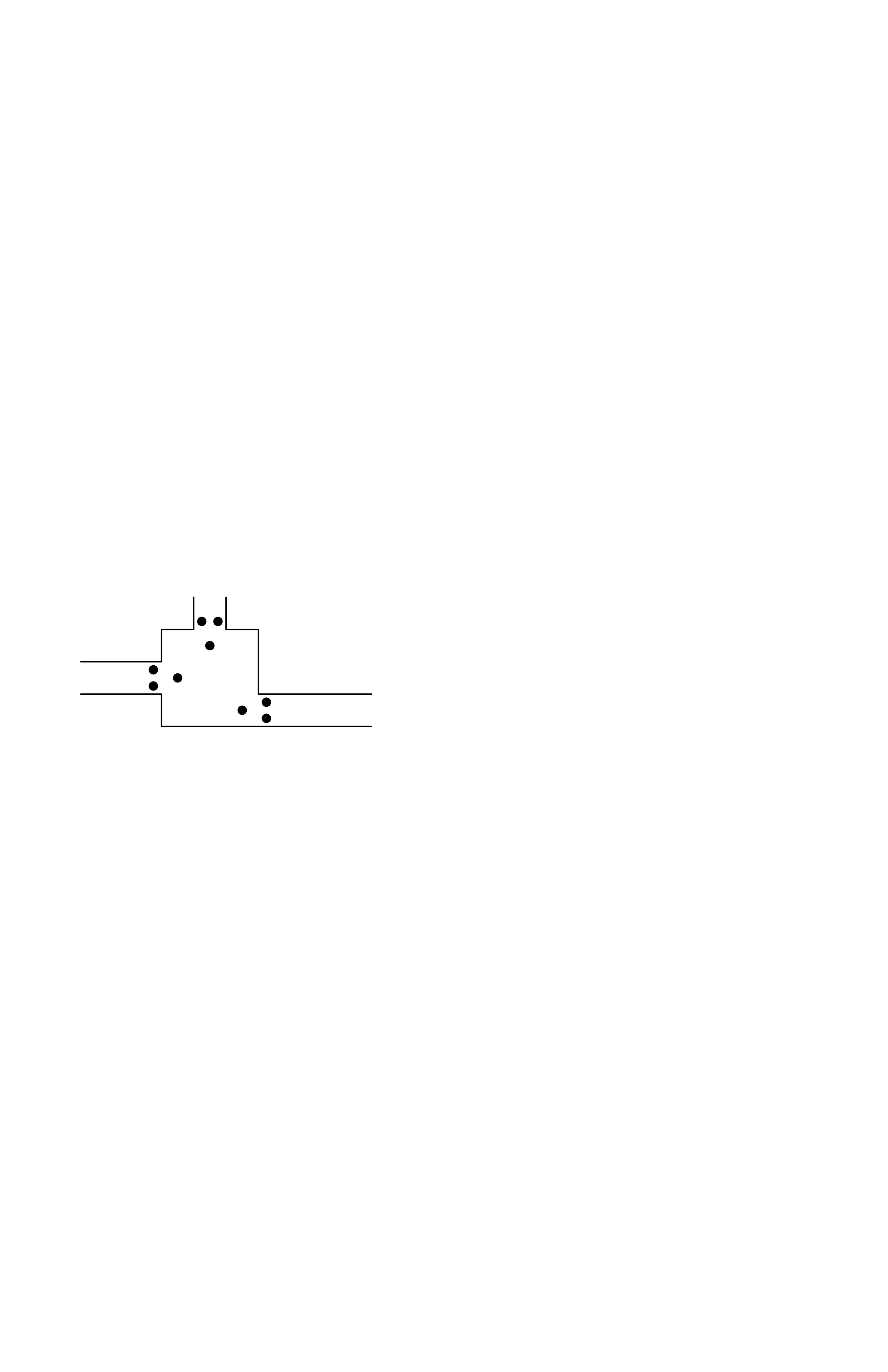} &
\hspace*{-5.25in}\includegraphics[scale=.5]{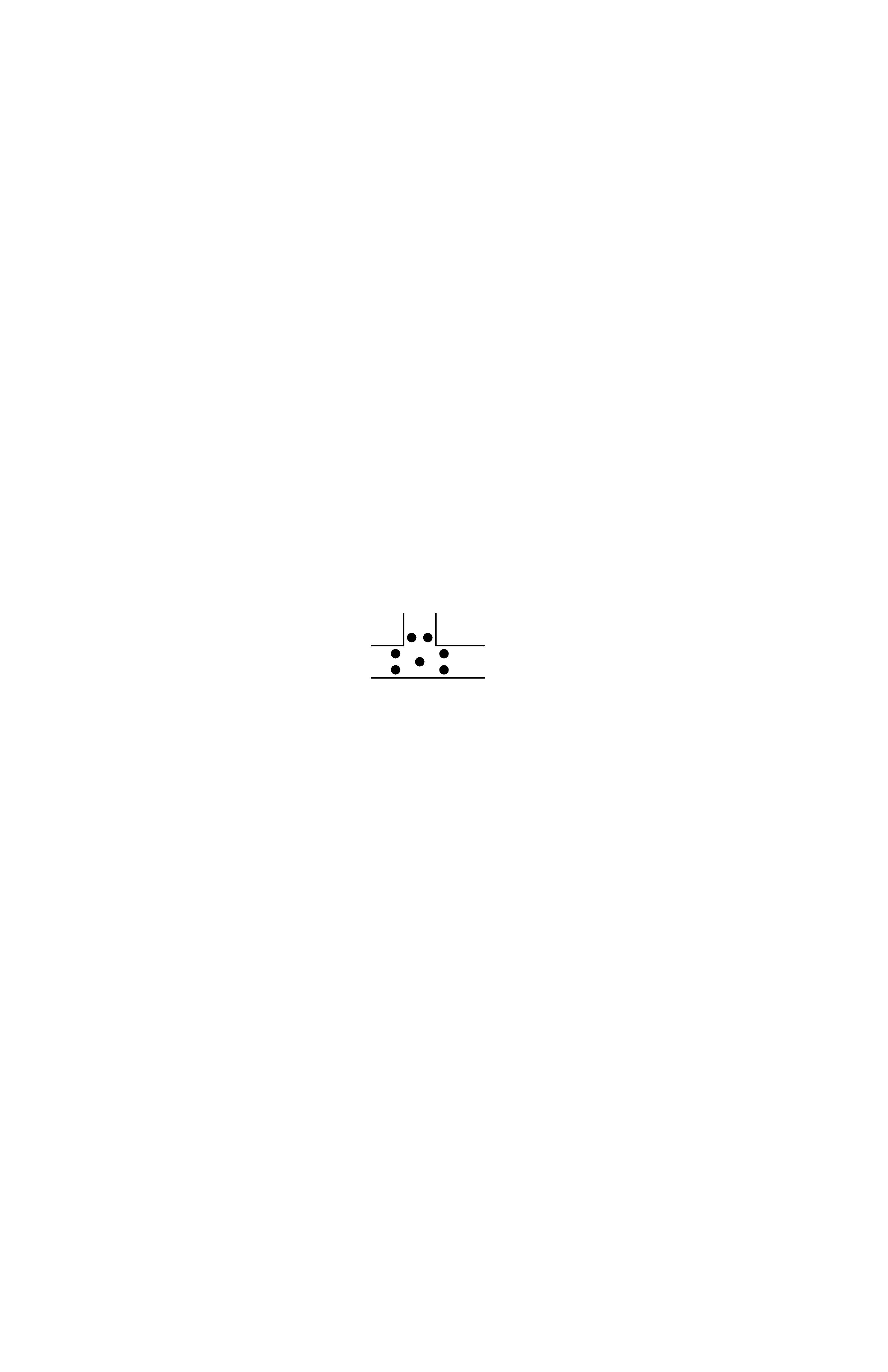} &
\hspace*{-4in}\includegraphics[scale=.5]{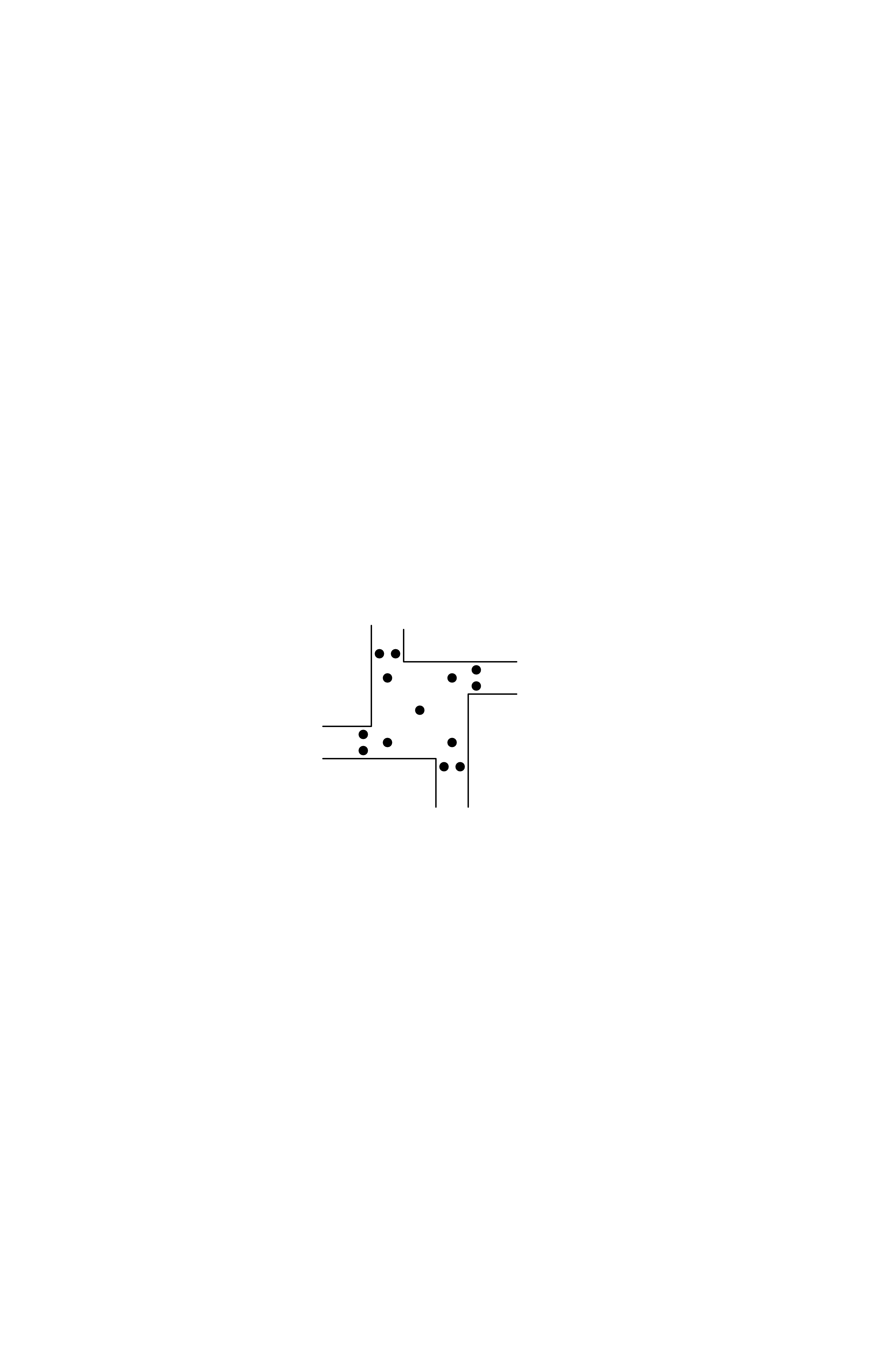} \\[-3.5in]
\hspace*{-3.2in}(a) & 
\hspace*{-5.2in}(b) &
\hspace*{-4in}(c)
\end{tabular} 
\caption{\label{fig-box-tuple} Gadgets for reducing 3DM to rectangular
$k$-anonymization in the plane.
(a) the tuple gadget; (b) the chooser gadget; (c) the cross-over gadget.}
\end{figure}

When we have completed this construction, we will have reduced 3DM to
a rectangle $k$-anonymization problem inside a rectilinear polygon $P$
containing holes that has its points and polygon vertices on a polynomial-sized integer grid. 

To complete the construction, then, we place $3$ (identical) points
at every grid location that is not properly in the interior of $P$.
Each such set of three points must be partitioned into a separate
rectangle, which will block any rectangle containing points 
properly inside $P$ from crossing the boundary of $P$ without
increasing its weight greater than $k$.
Thus, we can ``erase'' $P$ at this point and we will have reduced 3DM
to an instance of rectangular $k$-anonymization in the plane, for
$k=3$.

\subsection{An Approximation Algorithm for GPS Coordinates}
In this subsection, we provide an approximation algorithm
for GPS coordinates.
Suppose, therefore, that we are given a set of points $S$
and we wish to 
partition the plane into axis-aligned rectangles so as to minimize
the maximum weight of any rectangle.
We construct a kd-tree on $S$, using the cutting rule of always
splitting a rectangle with an axis-aligned cutting line if it is
possible to create two subrectangles each of weight at least $k$.
When we can no longer cut rectangles, satisfying this criterion, we
stop.
We note that this will produce a good approximation to the optimal
solution, with a worst-case 
degenerate case being four points of multiplicity $k-1$
placed at the N, S, E, and W directions of a point of multiplicity
$k$.
It may be possible for such a configuration to be partitioned into
rectangles of size $k$, whereas our approximation may, in the worst
case, produce a rectangle of weight $5k-4$ in this case.
Therefore, we have the following:
\fi

\label{sect:GPS}
\begin{theorem} \label{thm:GPS}
There is a polynomial-time 
approximation algorithm for rectangular generalization,
with respect to $k$-anonymization in the plane,
that achieves an approximation ratio of~5 in the worst case.
It is not possible for a polynomial-time algorithm to achieve an
approximation ratio better than $1.33$ unless P$=$NP.
\end{theorem}

\ifFullPaper
We note that a similar reduction to the one we give above
can be used to
show that no polynomial-time algorithm can achieve an approximation
ratio better than $1.33$ for the RTILE problem, unless P$=$NP, which
improves the previous lower bound for this problem of
$1.25$~\cite{kmp-oartp-98}.
\fi

\section{The Min-Max Bin Covering Problem}
In this section, we examine
single-attribute generalization,
with respect to the problem of $k$-anonymization for unordered data,
where quasi-identifying attribute values are arbitrary labels that
come from an unordered universe.  
(Note that if the labels were instead drawn from an ordered universe,
and we required the generalization groups to be intervals, the
resulting one-dimensional $k$-anonymization problem could be solved
optimally in polynomial time by a simple dynamic programming algorithm.) Our
optimization problem, then, is to generalize the input labels into
equivalence classes so as to minimize the maximum size of any
equivalence class, subject to the $k$-anonymization constraint.

It is convenient in this context
to use the terminology of bin packing; henceforth in
this section we refer to the input labels as {\sl items}, the
equivalence classes as {\sl bins}, and the entire generalization as
a {\sl packing}.  The \emph{size} of an item corresponds in this way to the number of records having a given label as their attribute value. Thus the problem becomes the
following, which we call the {\sl Min-Max Bin Covering Problem}:

\begin{list}{}{}
\item
Input: Positive integers $x_1,x_2,\ldots,x_n$  
and an integer nominal bin capacity $k>0$.
\item
Output: a partition of $\{1,2,\ldots,n\}$ into subsets $S_j$,
satisfying the constraint that, for each $j$, 
\begin{equation} \label{gl:constraint}
\sum_{i\in S_j} x_i \ge k,
\end{equation}
and minimizing the objective function 
\begin{equation} \label{gl:objective}
\max_j \sum_{i\in S_j} x_i.
\end{equation}
\end{list}

\newcommand{\cost}{\hbox{\sl cost}}
\noindent We will say that a partition satisfying (\ref{gl:constraint})
for all $j$ is {\sl feasible}, and the function shown in
(\ref{gl:objective}) is the {\cost} of this partition.  Note that
any feasible solution has cost at least $k$.

\subsection{Hardness Results}
In this subsection, we show that Min-Max Bin Covering
is NP-hard in the strong sense.  We begin by
converting the problem to a decision problem by adding a parameter
$K$, which is intended as an upper bound on the size of any
bin: rather than minimizing the maximum size of an
bin, we ask whether there exists a solution in which all
bins have size at most $K$.  This problem is clearly in NP.

We show that Min-Max Bin Covering is NP-hard
by a reduction from the following problem,
which is NP-complete in the strong sense~\cite{gj-cigtn-79}.

\begin{itemize}
\item
\sloppy
\textbf{3-Partition}.
Given a value $B$, and
a set $S$ of $3m$ weights $w_1,w_2,\ldots,w_{3m}$
each lying in $(B/4,B/2)$, such that
$\sum_{i=1}^{3m} w_i = m B,$
can we partition $\{1,2,\ldots,3m\}$ into sets $S_j$
such that for each $j$, $\sum_{i\in S_j} w_i = B$?
(Note that any such family of sets $S_j$ 
would have to have exactly $m$ members.)
\end{itemize}

For the reduction we simply let $x_i=w_i$ and $k=K=B$.
If the 3-Partition problem has answer yes,
then we can partition the items into $m$ sets each of total size
$k=K=B$ so the Min-Max Bin Covering problem has answer yes.
If, on the other hand, the 3-Partition problem has answer no,
no such partition is possible, so we have
\begin{theorem}
Min-Max Bin Covering is NP-complete in the strong sense.
\end{theorem}

In the preprint version of this paper~\cite{degl-k-09},
we show that there are limits on how well we can approximate
the optimum solution (unless P${}={}$NP):

\label{sect:gl:2approxlimit}
\begin{theorem} \label{gl:2approxlimit}
Assuming P${}\ne{}$NP,
there does not exist a polynomial-time algorithm
for Min-Max Bin-Covering
that guarantees an approximation ratio better than 2 (when inputs are expressed in binary),
or better than 4/3 (when inputs are expressed in unary).
\end{theorem}

\subsection{Achievable Approximation Ratios}

While the previous section shows that sufficiently small approximation ratios are
hard to achieve, in this section we show that we can establish larger approximation bounds
with polynomial time algorithms. The algorithms in this section can handle inputs that are expressed either in unary or binary, so they are governed by the stronger lower bound of 2 on the approximation ratio given in Theorem~\ref{gl:2approxlimit}. If $A$ is some algorithm for Min-Max Bin Covering Problem, and $I$
is some instance, let $A(I)$ denote the cost of the solution obtained
by $A$.  Let $\opt(I)$ denote the optimum cost for this instance.

Note that if
$\sum_{i=1}^n x_i < k$, there is no feasible solution; we will therefore
restrict our attention to instances for which
\begin{equation} \label{gl:assume.feasible}
\sum_{i=1}^n x_i \ge k.
\end{equation}

An approximation ratio of three is fairly easy to achieve.

\begin{theorem} \label{gl:approx3} Assuming {\rm (\ref{gl:assume.feasible})}
there is a linear-time algorithm $A$ guaranteeing that 
$$A(I) \le \max(k - 1 + \max_{i=1}^n x_i, 3k-3).$$
\end{theorem}

\proof Put all items of size $k$ or greater into their own bins, and
then, with new bins, 
use the Next Fit heuristic for bin covering (see \cite{AJKL:DualBP}) for
the remaining items, i.e.,
add the items one at a time, moving to a new bin once the current
bin is filled to a level of at least~$k$.
Then all but the last bin in this packing have level at most $2k-2$,
as they each have level
at most $k-1$ before the last item value is added and this last item
has size less than $k$. There may be one leftover bin with level less than
$k$ which must be merged with some other bin, leading to the claimed
bound.\qed

With a bit more effort we can improve the approximation ratio.
For convenience, in the remainder of this section we scale the problem by
dividing the item sizes by~$k$.  Thus
each bin must have level at least~1, and the item sizes are multiples
of $1/k$.

\newif\ifenoughspace \enoughspacefalse
\ifenoughspace
\begin{lemma}  \label{gl:2part}
Suppose we are given a list of positive
numbers $x_1,x_2,$ $\ldots,x_n$,
with each $x_i\le m$ and $\sum_{i=1}^n x_i = s$.
Then we can partition the list into two sublists
each having a sum of at most $(s+m)/2$.
\end{lemma}

\proof  Form the partial sums $p_i = \sum_{j=1}^i x_i$,
$i=0,1,\ldots,n$.  Then successive partial sums are clearly at most
$m$ apart, so we can find some $\ell$ so that $(s-m)/2 \le p_\ell \le
(s+m)/2$.  Hence if we cut the list in half just after the $\ell$th
element we will have two sublists summing to at most $(s+m)/2$.\qed
\fi

\begin{lemma} \label{gl:3part}
Suppose we are given a list of numbers $x_1,x_2,$ $\ldots,$ $x_n$,
with each $x_i\le 1/2$ and $\sum_{i=1}^n x_i = 1$.
Then we can partition the list into three parts
each having a sum of at most $1/2$.
\end{lemma}

\ifenoughspace
\proof
For convenience assume the $x_i$ appear in decreasing order.
Let the first sublist contain only $x_1$,
so its sum is clearly at most~1/2.
The remaining list has sum $1-x_1$ and no entries greater than
$x_1$, so by Lemma~\ref{gl:2part} we can partition it into two lists
each having sum at most $(1 - x_i + x_i)/2=1/2$.\qed
\else
\proof Omitted.
\fi

\begin{theorem} \label{gl:thm5/2}
There is a polynomial algorithm to solve Min-Max Bin Packing
with an approximation factor of $5/2$.
\end{theorem}

\proof  We will assume without loss of generality
that $\opt(I)\ge6/5$,
since otherwise the algorithm of Theorem~\ref{gl:approx3} could give
a $5/2$-approximation.

Assume the items are numbered in order of decreasing size.
Pack them greedily in this order into successive bins,
moving to a new bin when the current bin has level at least~1. Note that
then all of the bins will have levels less than~2,
and all of the bins except the last will
have level at least~1.  If the last bin also
has level at least~1, this packing is feasible
and has cost less than 2, so it is within a factor of 2 of the
optimum.

Next suppose that the last bin has level less than~1.  
We omit the details for the case in which we have formed at most 3 bins,
and subsequently we assume we have formed at least 4 bins.

Now let $f$ be size of the largest
item in the final bin, and let $r$ be the total size of the other items
in the last bin.  
Call an item {\sl oversize} if its size is at least~1,
{\sl large} if its size is in~$(1/2,1)$,
and {\sl small} if its size is at most $1/2$.
Consider two cases.

Case 1.  $f \le 1/2$.  
Then all items in the last bin are small,
so by Lemma~\ref{gl:3part} we can partition them into three sets,
each of total size at most 1/2.  Add each of these sets to one of the first
three bins, so no bin is filled to more than 5/2,
unless it was one of the bins containing an oversize item.
(We no longer use the last bin.)
Thus we have achieved an approximation ratio of 5/2.

Case 2.  $f > 1/2$.  
Note that in this case there must
be an odd number of large items, since each bin except the last
has either zero or exactly two large items.  Note also that $r$ in this case
is the total size of the small items, and $r<1/2$.
Let $x_1$ be the first large item packed.  If $x_1$ lies in the last
bin, we must have packed at least one oversize item.  Then moving
all of the items from the last bin (which will no longer be used) 
into the bin with this oversize item 
guarantees a 2-approximation.  Thus assume $x_1$ is not in the last bin.

Case 2.1.  $x_1 + r \ge 1$.  Then swap items $x_1$ and $f$, so the
last bin will be filled to a level $x_1 + r \in [1,2]$.
Also, the bin now containing $f$ will contain
two items of size in the range [1/2,1]
and thus have a level in the range [1,2].  Thus we have
a solution that meets the constraints and has cost at most 2.

Case 2.2.  $x_1 + r < 1$.  Since $r$ is the total size of the small
items, if any bin had only one large item it could not have level at
least 1 (as required for feasibility) and at most 6/5 (as required since
$\opt(I)\le6/5$).  Thus the optimum solution has
no bin containing only one large item.  Since there are an odd number
of large items, this means that the optimum solution has at least one
bin with 3 or more large items, so the cost of the optimum solution is
at least 3/2.  But then since the simple algorithm of
Theorem~\ref{gl:approx3} gives a solution of cost less than 3, it
provides a solution that is at most twice the optimum. \qed

\subsection{A Polynomial Time Approximation Guaranteeing a Ratio Approaching~2}

With more effort we can come arbitrarily close to the 
lower bound of 2 on the approximation factor given in
Theorem~\ref{gl:2approxlimit} for the binary case, with a
polynomial algorithm.

\begin{theorem} \label{gl:2pluseps}
For each fixed $\epsilon>0$, there is a polynomial time algorithm $A_\epsilon$
that,  given some instance $I$ of Min-Max Bin Covering,
finds a solution satisfying
\begin{equation} \label{gl:2+epsApprox}
A_\epsilon(I) \le (1+\epsilon)\bigl(\opt(I)+1\bigr).
\end{equation}
\end{theorem}

(The degree of the polynomial becomes quite large as $\epsilon$
becomes small.)

\proof 
The idea of the proof is similar to
many approximation algorithms for bin packing (see in particular
\cite[Chapter 9]{vazirani-aa});
for the current problem, we have to be especially
careful to ensure that the solution constructed is feasible.

We can assume that the optimum cost is at most 3, by the following
reasoning.  Say an item is {\sl nominal} if its size is less than~1,
and {\sl oversize} if its size is greater than or equal to~1.  First
suppose the total size of the nominal items is at least~1 and some
oversize item has size at least 3.  Then the greedy algorithm of
Theorem~\ref{gl:approx3} achieves an optimum solution, so we are done.
Next suppose the sum of the nominal items is at least~1 and no
oversize item has size at least 3.  Then the greedy algorithm of
Theorem~\ref{gl:approx3} achieves a solution of cost at most 3, so the
optimum is at most 3.  Finally suppose that the total size of the
nominal items is less than~1.  
Then there must be an optimum solution in which every
bin contains exactly one oversize item (and possibly some nominal items).  
Let $t_0$ (resp.~$t_1$)
be the size of the smallest (resp.~largest)
oversize item.  If $t_1-t_0\ge 1$, then we can form an optimum solution
by putting all nominal items in a bin with $t_0$.
If on the other hand $t_1 - t_0 < 1$,
we can reduce the
size of all oversize items by $t_0-1$ without changing the structure
of the problem, after which all oversize items will have size at most 2, and
the optimum will be at most 3.

\newcommand{\gltype}{\hbox{{\sl type}}} 
\newcommand{\glround}{{\hbox{\sl round}}}

Now call those items that have size greater than or equal to
$\epsilon$ {\sl large}, and the others {\sl small}.
Let $b=\sum_{i=1}^n x_i$; note that
$b\le 3n$, and any feasible partition will have at most $b$ bins.
Let $N$ be the largest integer for which $\epsilon(1+\epsilon)^N$
is less than three; note that $N$ is a constant depending only on $\epsilon$.
Let 
$$
\hat S = \bigl\{\epsilon(1+\epsilon)^\ell: \ell \in \{0,1,2,\ldots,N\}\bigr\}.
$$
For any item size $x$, define $\glround(x)$ to be the largest value in $\hat S$
that is less than or equal to $x$.
Let the {\sl type} of a packing $P$, written $\gltype(P)$, be the 
result of discarding all small items in $P$, and 
replacing each large $x_i$ by $\glround(x_i)$.
Note that any type can be viewed as
a partial packing in which the bins contain only items with sizes in $\hat S$.

Since, for fixed $\epsilon$, there are only a constant number of item
sizes in $\hat S$, and each of these is at least $\epsilon$, there are
only finitely many ways of packing a bin to a level of at
most 3 using the rounded values; call each of these ways a {\sl
configuration} of a bin.  Since the ordering of the bins does not
matter, we can represent the type of a packing by the number of times
it uses each configuration.  It is not hard to show that for fixed
$\epsilon$, as in the proof of \cite[Lemma 9.4]{vazirani-aa}, there
are only polynomially many types having at most $b$ bins.  (Of course,
for small $\epsilon$, this will be a polynomial of very high degree.)
We will allow types that leave some of the bins empty, allowing them
to be filled later. 

The algorithm proceeds as follows.  Enumerate all possible types $T$
that can be formed using the rounded large item sizes.
For each such type $T$ carry out the following steps:
\begin{enumerate}
\item Let $T'$ be the result of replacing each item $x$ in $T$,
which resulted from rounding some original input item $x_i$,
by any one of the original items $x_j$ such that $x=\glround(x_j)$,
in such a way that the set of items in $T'$ is the same as the
set of large items in the original input.
Note that there is no guarantee that $x_i=x_j$, since the rounding
process does not maintain the distinct identities of different items that 
round to the same value in $\hat S$.  However, we do know that
$\glround(x_i)=\glround(x_j)$, so we can conclude that
$x_j/x_i\in \bigl((1+\epsilon)^{-1},1+\epsilon\bigr)$.
\item Pack the small items into $T'$ by processing them in an arbitrary
order, placing each into the bin with the lowest current level.
Call this the {\sl greedy completion}~of~$T$.
\item Finally, while any bin has a level less than 1,
merge two bins with the lowest current levels. 
Note that this will lead to a feasible packing
because of (\ref{gl:assume.feasible}).
Call the resulting packing ${\cal F}(T)$, and let
$\cost\bigl({\cal F}(T)\bigr)$ be the maximum level
to which any bin is filled.
\end{enumerate}
Return the packing ${\cal F}(T)$ that minimizes
$\cost\bigl({\cal F}(T)\bigr)$ over all $T$.

We now show that (\ref{gl:2+epsApprox}) holds.
Let $P^*$ be a feasible packing achieving $\opt(I)$,
and let $P^*_{\rm large}$ be the result of discarding the small items in
$P^*$ (retaining any bins that become empty).
Consider the type $T$ obtained by rounding
all large items in $P^*_{\rm large}$
down to a size in $\hat S$.
Note that this must be one of the 
types, say~$T$, considered in the algorithm.
When we perform step~1 on~$T$, we obtain 
a packing $T'$ such that 
$\cost(T')\le (1+\epsilon)\cost(P^*)$.

If any level in the greedy completion is greater than
$(1+\epsilon)\opt(I)+\epsilon$,
then during the greedy completion all bins must have reached a level
greater than $(1+\epsilon)\opt(I)$, so their total size would
be greater than $(1+\epsilon)\sum_{i=1}^n x_i$, 
contradicting the fact that the greedy completion
uses each of the original items exactly once.
Thus all bins in the greedy completion
have level at most $(1+\epsilon)\opt(I)+\epsilon$.
Also, it cannot be that all bins in the greedy completion have level
less than~1, since then the total size of the items would be
less than the number of bins, contradicting the fact that the optimum
solution covers all the bins.

During step~3, as long as at least two bins have levels
below~1, two of them will be merged to form a bin with a level at
most~2.  If then only one bin remains with a level below~1,
it will be merged with a bin with level in
$\bigl[1,(1+\epsilon)\opt(I)+\epsilon\bigr)$
to form a feasible packing with no
bin filled to a level beyond $(1+\epsilon)\opt(I)+1+\epsilon$, as
desired.\qed

Note that the bound of Theorem~\ref{gl:2pluseps} implies
$A_\epsilon(I) \le 2(1+\epsilon)\opt(I)$.

We also note that if one is willing to relax both the feasibility constraints
and the cost of the solution obtained, a polynomial-time $(1+\epsilon)$
approximation scheme of sorts is possible.  (Of course, this would not
guarantee $k$-anonymity.)
\begin{theorem} \label{gl:1pluseps}
Assume that all item sizes $x_i$ in the input are expressed in binary,
and let $\epsilon > 0$ be a fixed constant.
There is a polynomial time algorithm that, 
given some instance $I$ of Min-Max Bin Covering,
finds a partition of the items into disjoint bins $S_j$
such that
\begin{equation} \label{gl:modconstraintobjective}
\forall j ~ \sum_{i\in S_j} x_i \ge 1 - \epsilon,
{~~~~~\rm and~~~~~}
\max_j \sum_{i\in S_j} x_i \le (1 + \epsilon) \opt(I).
\end{equation}
\end{theorem}

\proof[sketch]  Roughly, one can use an algorithm similar to
that of the previous theorem
but omitting the last phase in which we merge bins to eliminate
infeasibility.  We omit the details.\qed

\ifFullPaper
\section{Experimental Results}
In this section, we give experimental results for
implementations and extensions of a couple of our approximation algorithms.
Because of space limitations, we focus here on approximate
$k$-anonymization for unordered data.

So as to represent a real distribution of quasi-identifying information,
we have chosen to use the following data sets provided by the
U.S.~Census Bureau from the 1990 U.S.~Census:
\begin{itemize}
\item
\textbf{FEMALE-1990}:
Female first names and their frequencies, for names with frequency at
least 0.001\%.
\item
\textbf{MALE-1990}:
Male first names and their frequencies, for names with frequency at
least 0.001\%.
\item
\textbf{LAST-1990}:
Surnames and their frequencies, for surnames with frequency at
least 0.001\%.
\end{itemize}

For each data set, we ran a number of experiments, so as to test the
quality of the approximations produced by the method of 
Theorem~\ref{gl:approx3}, which we call \emph{Fold},
and compare that with the quality of the approximations produced by a
simplified version of the method of Theorem~\ref{gl:thm5/2}, 
which we call \emph{Spread}, for all
values of $k$ ranging from the frequency of the most common name 
to the value of $k$ that results in there being only two
equivalence classes.

The simplification 
we implemented for the method of Theorem~\ref{gl:thm5/2}
involves the distribution of left-over items
at the end of the algorithm.
In this case, we distribute left-over items
among existing equivalence classes using a greedy algorithm, where we
first add items to classes that have less than the current maximum
until adding an item to any class would increase the maximum.
At that point, we then distribute the remaining items to equivalence
classes in a round-robin fashion.

We tested both approaches on each of the above data sets,
with the data being either randomly
ordered or sorted by frequencies. 
For each test, we analyzed the ratio of the size
of the largest equivalence class to $k$, the anonymization parameter,
since this ratio serves as an upper bound on the algorithm's
approximation factor.
The overfull ratios for each algorithm is reported for each of the
above data sets in Figure~\ref{fig:names}.

\begin{figure}[hbt]
\centering

~
\vspace{-0.75in}

\includegraphics[width=4in]{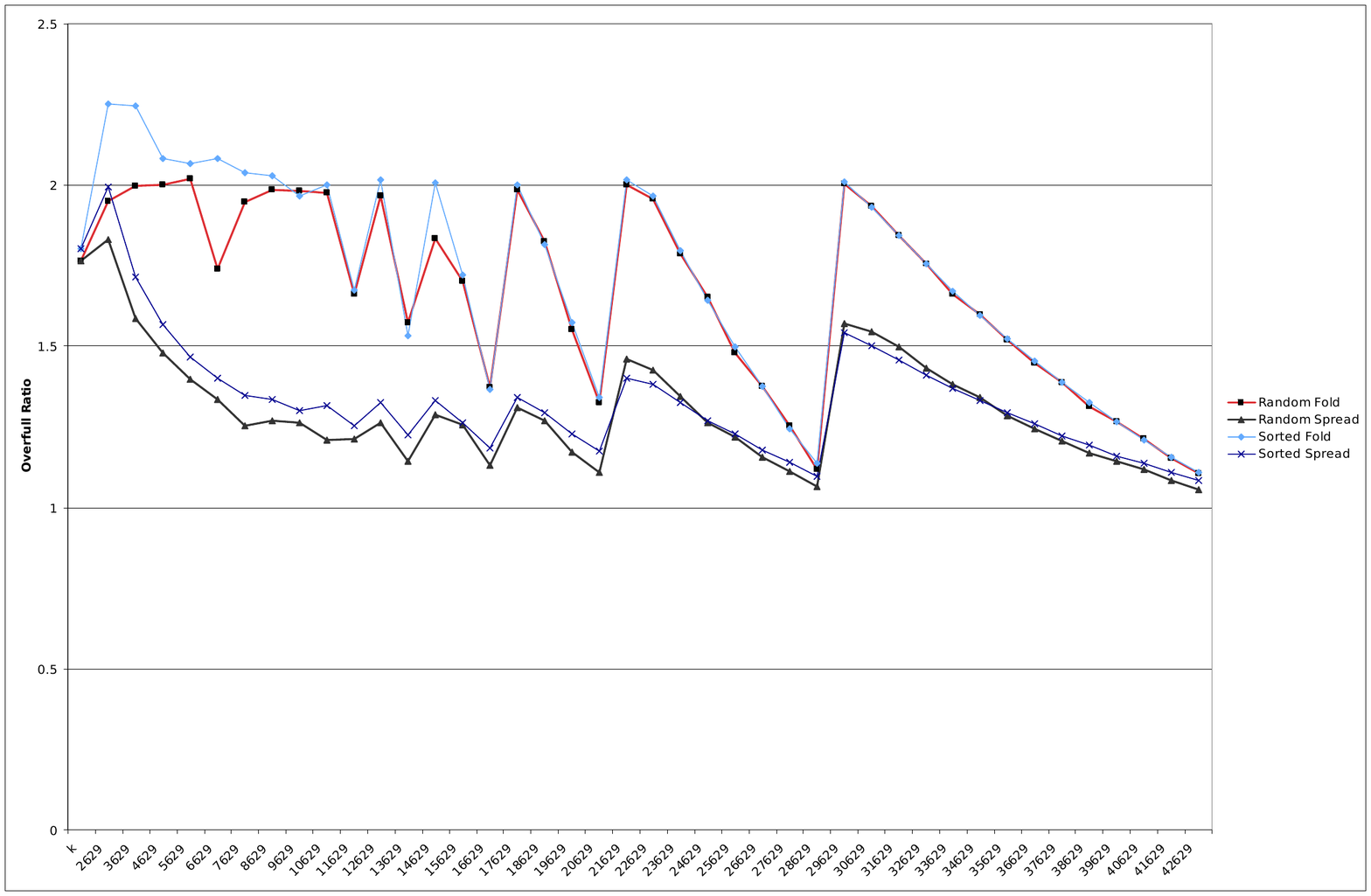}

\vspace*{-0.9in}

\includegraphics[width=4in]{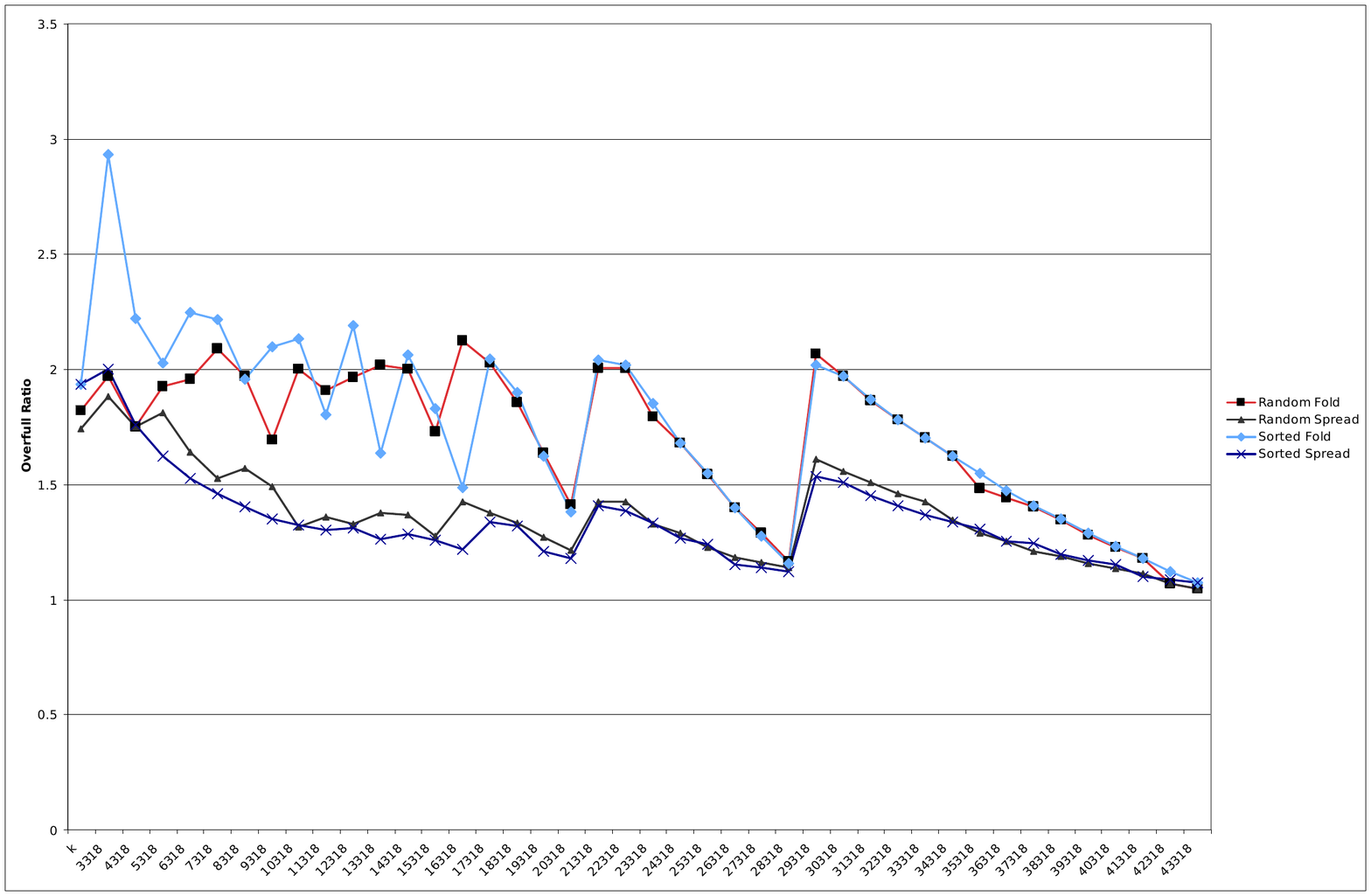}

\vspace*{-0.9in}

\includegraphics[width=4in]{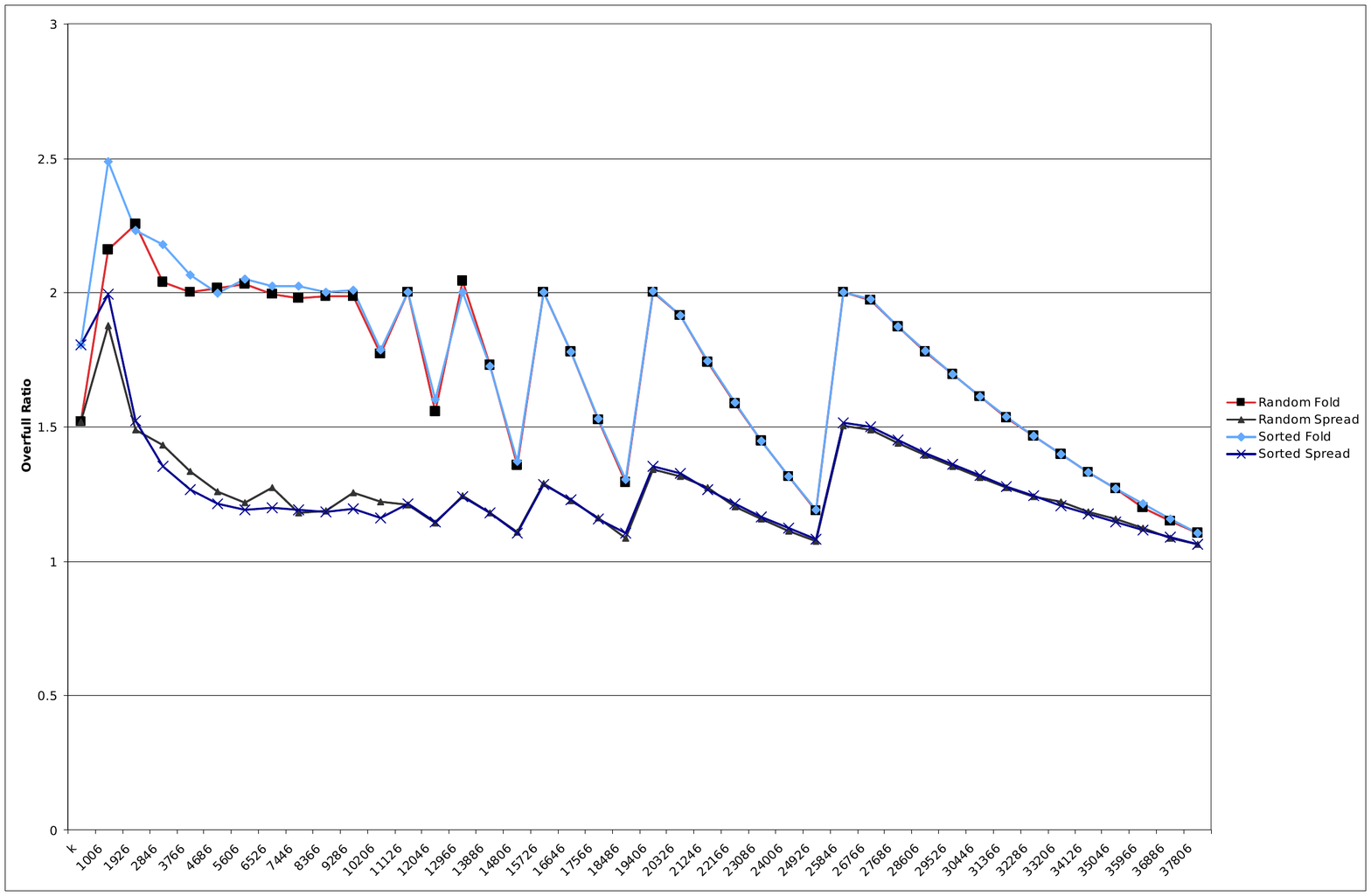}

\vspace*{-0.6in}

\caption{Overfull ratios for the FEMALE-1990, MALE-1990, and LAST-1990
data sets, respectively.
Maximum ratios are reported
for each subrange with respect to four algorithms: 
Random Fold, which is the Fold algorithm on randomly-ordered data,
Random Spread, which is the Spread algorithm on randomly-ordered data,
Sorted Fold, which is the Fold algorithm on ordered data,
Sorted Spread, which is the Spread algorithm on ordered data.
}
\label{fig:names}
\end{figure}

There are number of interesting observations we can make from our
experimental results, including the following:
\begin{itemize}
\item
The Spread algorithm is superior to the Fold algorithm, for both
randomly-ordered and sorted data.
\item
Generalizing data into equivalence classes based on a random ordering
of the frequencies is often superior to a sorted order.
\item
When considering values of $k$ in increasing order,
there are certain threshold values of the parameter $k$ where the
number of equivalence classes drops by one, and this drop has a
negative effect on the overfull ratio. The negative effect is
especially pronounced for the Fold algorithm.
(This behavior is what causes the increasing ``jagginess'' towards the right
of the ratio plots.)
\item
The performance of both the Fold and Spread algorithms on these
real-world data sets is much better than 
the worst-case analysis.
\end{itemize}

Thus, our algorithms confirm 
our intuition about the Spread algorithm being better than the Fold
algorithm. In addition, our experimental analysis shows that the Spread
algorithm performs quite well on real-world data sets.

\fi
\ifFullPaper
\section{Future Directions}
There are a number of interesting directions for future work. For
example, real world data sets often have two or three
quasi-identifying attributes (such as zip-codes and disease name
labels). Our results show that $k$-anonymization problems in such
cases are NP-hard, but there are a host of open problems relating to
how well such multi-attribute 
problems can be solved approximately in polynomial time.

\fi

\paragraph{Acknowledgments.}
We would like to thank Padhraic Smyth and Michael Nelson 
for several helpful discussions related to the topics of this paper. 
This research was supported in part by the NSF under grants 0830403,
0847968, and 0713046.

{\small
\bibliographystyle{abbrv}
\bibliography{k_anonymity,du,data_mining}
}

\end{document}